\documentclass[aps,rmp,reprint,superscriptaddress,noeprint,nolongbibliography]{revtex4-2}

\usepackage[english]{babel}
\usepackage{graphics}
\usepackage{array}
\usepackage{graphicx}
\usepackage{todonotes}
\usepackage{amsfonts,amssymb,amsmath}
\usepackage{color}
\usepackage{threeparttablex}
\usepackage{longtable} 
\usepackage{physics}
\usepackage{comment}
\usepackage{braket}

\usepackage[separate-uncertainty,per-mode=fraction]{siunitx}
\usepackage[extra]{optional} 

\definecolor{darkblue}{rgb}{0.0,0.0,0.4}
\definecolor{darkgreen}{rgb}{0.0,0.4,0.0}
\definecolor{darkred}{rgb}{0.6,0.0,0.0}
\usepackage[bookmarksopen]{hyperref}
\hypersetup{colorlinks,linkcolor=darkblue,citecolor=darkblue,urlcolor=darkblue}

\renewcommand{\vector}[1]{\boldsymbol{#1}}
\newcommand{\cred}[1]{{#1}}

\newcolumntype{M}[1]{>{\centering\arraybackslash}m{#1}}

\renewcommand{\selectlanguage}[1]{}

\newcommand\textcomma{\textrm{, }}

\begin{document}
\title{Strongly dipolar molecular Bose-Einstein condensates: From few- to many-body physics}

\author{Andreas Schindewolf}
\affiliation{Vienna Center for Quantum Science and Technology\textcomma Atominstitut\textcomma TU Wien\textcomma  Stadionallee 2\textcomma  A-1020 Vienna\textcomma Austria}

\author{Jens Hertkorn}
\affiliation{MIT-Harvard Center for Ultracold Atoms\textcomma Research Laboratory of Electronics\textcomma and Department of Physics, Massachusetts Institute of Technology\textcomma  Cambridge\textcomma  Massachusetts 02139\textcomma USA}
\affiliation{5. Physikalisches  Institut  and  Center  for  Integrated  Quantum  Science  and  Technology, Universit\"at  Stuttgart\textcomma  Pfaffenwaldring  57\textcomma  70569  Stuttgart\textcomma  Germany}

\author{Ian Stevenson}
\affiliation{Department of Physics\textcomma Columbia University\textcomma New York\textcomma New York 10027\textcomma USA}

\author{Matteo Ciardi}
\affiliation{Institute for Theoretical Physics\textcomma  TU Wien\textcomma  Wiedner Hauptstraße 8-10/136\textcomma 1040 Vienna\textcomma Austria}

\author{Phillip Gross}
\affiliation{Vienna Center for Quantum Science and Technology\textcomma Atominstitut\textcomma TU Wien\textcomma  Stadionallee 2\textcomma  A-1020 Vienna\textcomma Austria}

\author{Dajun Wang}
\affiliation{Department of Physics\textcomma The Chinese University of Hong Kong\textcomma Shatin\textcomma Hong Kong SAR\textcomma China}

\author{\\Tijs Karman}
\affiliation{Institute for Molecules and Materials\textcomma Radboud University\textcomma Nijmegen\textcomma The Netherlands}

\author{Goulven Qu{\'e}m{\'e}ner}
\affiliation{Universit\'{e} Paris-Saclay\textcomma CNRS\textcomma Laboratoire Aim\'{e} Cotton\textcomma 91405 Orsay\textcomma France}

\author{Sebastian Will}
\affiliation{Department of Physics\textcomma Columbia University\textcomma New York\textcomma New York 10027\textcomma USA}

\author{Thomas Pohl}
\affiliation{Institute for Theoretical Physics\textcomma  TU Wien\textcomma  Wiedner Hauptstraße 8-10/136\textcomma 1040 Vienna\textcomma Austria}

\author{Tim Langen} 
\email{tim.langen@tuwien.ac.at}
\affiliation{Vienna Center for Quantum Science and Technology\textcomma Atominstitut\textcomma TU Wien\textcomma  Stadionallee 2\textcomma  A-1020 Vienna\textcomma Austria}
\affiliation{5. Physikalisches  Institut  and  Center  for  Integrated  Quantum  Science  and  Technology, Universit\"at  Stuttgart\textcomma  Pfaffenwaldring  57\textcomma  70569  Stuttgart\textcomma  Germany}

\begin{abstract}
Recent advances in molecular cooling have enabled the realization of strongly dipolar Bose--Einstein condensates (BECs) of molecules, and BECs of many different molecular species may become experimentally accessible in the near future. Here, we explore the unique properties of such BECs and the new insights they may offer into dipolar quantum fluids and many-body physics. We explore which parameter regimes can realistically be achieved using currently available experimental techniques, discuss how to implement these techniques, and outline which molecular species are particularly well suited to explore exotic new states of matter. We further determine how state-of-the-art beyond mean-field theories, originally developed for weakly dipolar magnetic gases, can be pushed to their limits and beyond, and what other long-standing questions in the field of dipolar physics may realistically come within reach using molecular systems.
\end{abstract}

\maketitle

\tableofcontents

\section{Introduction}\label{sec:intro}

\cred{Systems of dipolar particles play a distinctive role in quantum many-body physics due to the long-range and anisotropic character of their interactions. The anisotropy of the dipole--dipole interaction — whose sign and magnitude depend on the relative orientation of the dipoles — combined with the tunability afforded by external fields and geometric confinement, enables a degree of experimental control over the interaction potential that is rarely available. These properties can profoundly alter the spreading of entanglement and scrambling of quantum information \cite{Hauke2013,FossFeig2015,Chen2019,Tran2021}, affect transport and localization phenomena \cite{Yao2014}, and enable the exploration of exotic quantum magnetism and phases of matter in and out of equilibrium \cite{Koffel2012,Yao2018,Carroll2025}. In ultracold quantum gases, the realization of dipolar interactions has yielded numerous discoveries that deepened our understanding of correlations and quantum fluctuations in bosonic systems and enabled explorations of long-elusive quantum many-body states~\cite{Lahaye2009,Boettcher2021,Chomaz2023,Baranov2012,Bohn2017}.
Great advances in the study of atomic dipolar gases have been achieved using Bose--Einstein condensates (BECs) of different species of magnetic atoms~\cite{Griesmaier2005,Lu2011,Lu2012,Aikawa2012,Miyazawa2022}. Most notably, such systems have enabled the observation of dipolar quantum droplets and supersolids~\cite{Boettcher2021,Chomaz2019,Tanzi2019}, and they are predicted to show fascinating pattern-formation phenomena yet to be explored~\cite{Hertkorn2021,Zhang2021}. These possibilities provide a unique laboratory platform for broad studies of exotic phases of matter. Quantum droplets are a rare example of self-bound quantum matter that remains stable in free space without external confinement~\cite{Baillie2016,Schmitt2016,Chomaz2016}. They emerge from the attractive part of dipole--dipole interactions and are stabilized by quantum fluctuations, somewhat akin to the stabilization of nuclei or neutron stars~\cite{Poli2023}. Supersolids, by contrast, simultaneously exhibit the crystalline density order of a solid and the frictionless transport of a superfluid~\cite{Boninsegni2012}. The possibility of such a peculiar state was first suggested in the context of low-temperature helium more than 
50 years ago~\cite{Andreev1969,Leggett70,Chester1970, Gross58}, but evaded experimental verification there~\cite{Balibar2010}. Its realization in dipolar BECs has enabled thorough experimental investigations of supersolidity in recent years~\cite{Guo2019,Tanzi2019b,Natale2019}. 

Dipolar interactions between highly magnetic atoms have also been used to study thermalization~\cite{Tang2018}, to explore the physics of extended Bose--Hubbard models~\cite{Baier2016,Fraxanet2022,Su2023}, to investigate spin dynamics in synthetic quantum magnets~\cite{Lepoutre2019}, and to demonstrate protocols for quantum-enhanced sensing of magnetic fields~\cite{Chalopin2018}.}

In parallel to these developments, important progress has been made in the cooling of even more strongly dipolar molecules~\cite{Bohn2017,Langen2023}. Following decades of efforts, this has resulted in the creation of the first degenerate Fermi gases of ground-state molecules~\cite{DeMarco2019,Valtolina2020,Schindewolf2022,Duda2023,Cao2023}, directly laser-cooled molecules with significant phase-space density~\cite{Anderegg2018,Wu2021,Jurgilas2021,Jorapur2024,Burau2023}, near-degenerate assembled bosonic gases~\cite{Bigagli2023,Son2020,Lin2023} and, most recently, the first dipolar BECs and droplets of molecules~\cite{Bigagli2024,Shi2025DAMOP,Zhang2025b}. 

Taken together, these recent landmark advances raise the important question: How will dipolar physics affect a BEC of polar molecules—and, conversely, what novel insights might BECs of different molecular species, with different individual properties, offer to the study of dipolar physics in the near future?

Although conceptually similar to magnetic atoms, and historically also treated on the same theoretical footing~\cite{Lahaye2009}, molecules, with their substantially larger electric dipole moments, present unique challenges and opportunities when it comes to dipolar many-body physics. It is often assumed that molecules provide a straight-forward route to strongly-interacting dipolar systems. However, realizing this potential in practice entails subtle  challenges and limitations, which we will examine in detail. On the few-body side, significant experimental efforts are required to control long-range interactions and short-range losses~\cite{Miranda2011,Matsuda2020,Li2021,Valtolina2020,Anderegg2021,Schindewolf2022,Bause2023,Chen2023,Bigagli2023,Lin2023,Karam2023}, 
which influence achievable parameter regimes. On the many-body side, the important role that quantum fluctuations play in dipolar systems~\cite{Boettcher2021} and their connection to the limits of mean-field theory for molecular systems have so far only rarely been considered~\cite{Schmidt2022,Cardinale2025,Polterauer2025}.

Here, we address these open questions and explore them from both theoretical and experimental perspectives. Building on the established and well-tested theoretical framework to describe atomic dipolar quantum gases, we discuss differences and extensions for molecules, outline the principles that govern the characteristics of different molecular species, and explore the various existing collisional shielding mechanisms that are essential to the formation of stable ultracold molecular gases. We also discuss the experimental challenges in achieving ultracold molecular samples. We then explore known limitations in the theoretical framework of dipolar physics to highlight areas where molecules can provide a new perspective, go beyond existing mean-field approaches, benchmark more advanced theoretical models, and realize hitherto unexplored states of matter. 

\section{Interactions and few-body physics}
\label{sec:few-body}

We begin by examining the nature of interactions in ultracold molecular gases. We trace their evolution from fundamental few-body scattering processes to emergent many-body collective behavior.

\subsection{Long-range interactions}
\label{sub:sec:long-range}
In the literature, many references exist on the treatment of two-body interactions and collisions between ultracold dipolar magnetic atoms and electrically polar molecules, such as
\textcite{Napolitano1997,Avdeenkov2002,Krems2004,Micheli2007,Tscherbul2009,Lahaye2009,Baranov2012,Lepers2013,Wang2015,Quemener2017,Lepers2017,Karman2022,Chomaz2023}, as well as their connection to ultracold chemistry \cite{Liu2022,Karman2024}.

The interaction potential between two molecules can generically be written as the superposition of different power law contributions of the form
\begin{equation}
V_n({\bf r})=\frac{C_n}{r^n} f(\theta,\varphi), \label{eq:genericpotential}
\end{equation}
where $C_n$ parameterizes the strength of the respective contributions and the dimensionless function $f_n(\theta,\varphi)$ captures their angular dependence. In the following, we start by providing a general  overview of dipole--dipole interactions (DDIs) and their properties, which dominate the interaction between molecules at long range. 

For two generic dipoles oriented along unit vectors $\hat{\vector{e}}_1$ and $\hat{\vector{e}}_2$, \cred{the} DDI is described by
\begin{equation}
    V_\text{dd}^{\text{gen}}(\vector{r}) = \frac{C_3}{r^3} \left[ (\hat{\vector{e}}_1 \cdot \hat{\vector{e}}_2) - 3 (\hat{\vector{e}}_1 \cdot \hat{\vector{e}}_r) (\hat{\vector{e}}_2 \cdot \hat{\vector{e}}_r)\right]
\label{eq:Vddgeneral}
\end{equation}
where $\vector{r}$, often expressed in spherical coordinates $(r, \theta, \varphi)$, is the vector between the two centers of mass of dipoles 1 and 2, and $\hat{\vector{e}}_r = \vector{r}/r$ is the corresponding unit vector. The quantity $C_3=C_\text{dd} / 4 \pi$ characterizes the strength of the DDI.
For magnetically dipolar atoms, we have $C_\text{dd} = \mu_0\mu_\text{m}^2$ where $\mu_0$ is the the vacuum permeability and $\mu_\text{m}$ the magnitude of the atomic magnetic dipole moment. For electrically dipolar molecules,  $C_\text{dd} = d^2/\epsilon_0$, where $\epsilon_0$ is the vacuum permittivity and $d$ the molecular electric dipole moment. 

If we now consider two dipoles that are oriented with respect to an external field along a unit vector $\hat{\vector{e}}_z$ taken as the quantization axis, Eq.~\eqref{eq:Vddgeneral} reduces
to
\begin{equation}
    V_\text{dd}(\vector{r}) = \frac{C_3}{r^3}\left(1-3\cos^2{\theta}\right).
\label{eq:Vdd}
\end{equation}
Here, $\theta$ denotes the angle between the quantization axis and $\vector{r}$, highlighting the anisotropic and long-range nature of the DDI. 

The strength of DDI can be quantified by the \textit{dipolar length}, similar to how the scattering length reflects the strength of short-range contact interactions. Various definitions of the dipolar length exist in the literature. In the context of scattering theory \cite{Bohn2009} it is often defined as $a_\text{d} = C_\text{dd} \mu/(4\pi\hbar^2)$, where $\mu$ denotes the reduced mass.
Another common definition is $\tilde{a}_\text{d} = 2 a_\text{d}$ \cite{Gao2010}. 
In the context of dipolar many-body physics, $a_\text{dd} = C_\text{dd} M/(12\pi\hbar^2) = (2/3) a_d$, with the regular mass $M$, is defined such that the relative dipolar strength $\varepsilon_\text{dd} = a_\text{dd}/a_s$ marks the instability of a homogeneous BEC against collapse for $\varepsilon_\text{dd} \geq 1$ \cite{Chomaz2023,Lahaye2009}. 

The corresponding characteristic interaction energy scale is set by the dipolar energy $E_\text{d} = C_\text{dd}/(4\pi a_\text{d}^3) = (\hbar^6/\mu^3)(4\pi/C_\text{dd})^2$ or $\tilde{E}_\text{d} = E_\text{d}/8$, which notably decreases as $C_\text{dd}$ grows. 
This energy scale separates the threshold regime at low energies and the semi-classical regime at higher energies~\cite{Bohn2009}. In the semi-classical regime, the elastic scattering cross section is temperature dependent and can be approximated as
\begin{equation}
    \sigma_\text{scl} \approx \frac{8\pi a_\text{d}}{3k},
\end{equation}
with the thermally-averaged collisional wave-vector $k = \sqrt{\pi M k_\text{B} T}/\hbar$ \cite{Bohn2009,Bigagli2023}. 
In the threshold regime,
the cross section becomes energy independent and for indistinguishable bosons is
\begin{equation}
    \sigma_\text{thr} \approx \frac{32\pi}{45} a_\text{d}^2 +  8\pi a_s^2.
\end{equation}
The dependence of the elastic cross section on the direction of incidence, relevant for cross-dimensional thermalization during evaporative cooling and for other out-of-equilibrium dynamics \cite{Aikawa2014,Valtolina2020,Schindewolf2022,Wang2022,Wang2023,Chen2023,Bigagli2023,Lin2023,Wang2024b,Wang2025}, was for the semi-classical regime described by \textcite{Bigagli2023,Lin2023,Wang2024} and for the threshold regime calculated by \textcite{Bohn2014,Sykes2015,Wang2021}.

\subsection{Inducing dipoles in molecules}
\label{sub:sec:induced_dipoles}
In the absence of external fields, gases of both magnetic atoms and molecules feature no laboratory-frame dipole moment. In this review, we are concerned with dipolar molecules, which---unlike atoms---possess a permanent (i.e., intrinsic) molecule-frame electric dipole moment, labeled $d_0$ in their vibrational ground state. A laboratory-frame dipole can be induced by applying a static electric field. Alternatively, oscillating electromagnetic fields, such as microwaves (MW), can coherently couple internal states to induce dipole moments. 

Specifically, the induced laboratory-frame dipole moments are a result of coupling rotational states of opposing parity. For clarity, we denote rotational states as $\lvert N, m_N \rangle$, where $N$ is the rotational quantum number and $m_N$ its projection onto the quantization axis $\hat{\vector{e}}_z$. In cases where electronic-spin degrees of freedom are relevant, one often also uses the coupled angular momentum quantum number $J$ and projection $m_J$. Both notations appear in the literature, and we will employ them accordingly whenever they are most suitable.

\paragraph{Static fields}
More specifically, to realize \cred{dipoles oriented in the lab frame}, as described in Eq.~\eqref{eq:Vdd}, a static electric field $\vector{E} = E \, \hat{\vector{e}}_z$ can be applied to the molecules \cite{Quemener2011, Lassabliere2022}. The field polarizes the molecules by hybridizing their bare rotational states. As a result, dressed rotational states $\ket{\tilde{N}, m_N}$ are established, which carry a finite dipole moment along the field direction. This induced dipole moment is defined as the expectation value of the electric dipole moment along the quantization axis 
$d_\text{ind} = \bra{\tilde{N}, m_N} \vector{d}_0 \cdot \vector{e_z} \ket{\tilde{N}, m_N}$. Here, $C_\text{dd} = d_\text{ind}^2/\epsilon_0$ where $d_\text{ind} < d_0$ is now the induced dipole moment along the quantization axis.

Generally, the calculation of the induced dipole moment requires a full diagonalization of the Stark Hamiltonian. A straightforward approximation can be applied to Hund's case (b) states, which are commonly encountered in assembled bialkali molecules. If the static electric field is weak enough so that $d_0 E \ll B_\text{rot}$ ($ B_\text{rot}$ is the rotational constant of the molecule), it is found that $d_\text{ind} = \frac{d_0^2 E}{B_\text{rot}} \frac{3m_N^2/[N(N+1)]-1}{(2N-1)(2N+3)}$ \cite{Micheli2007}, for a given rotational state $\ket{\tilde{N}, m_N}$. 
Typically, $d_\text{ind} = d_0^2 E / 3 B_\text{rot}$ and $d_\text{ind} = - d_0^2 E / 5 B_\text{rot}$ 
for the two lowest rotational states $\ket{\tilde{0}, 0}$ and $\ket{\tilde{1}, 0}$.

\paragraph{Microwave fields}
The interactions in MW-dressed molecules can significantly extend these simple bare dipolar interactions beyond those described by Eqs.~\eqref{eq:Vddgeneral} and \eqref{eq:Vdd}, and \cred{introduce additional tunable features in the collisional processes.}

\begin{figure}
    \centering
    \includegraphics[width=0.95\columnwidth]{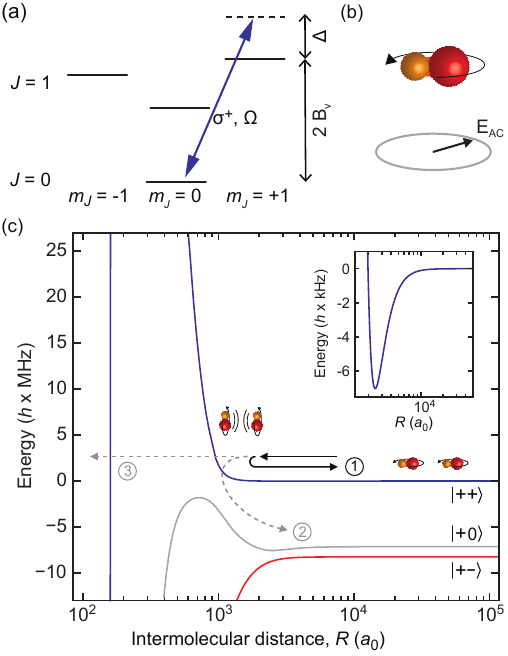}
    \caption{ Microwave shielding of a $^1\Sigma^+$ molecule. (a) Lowest rotational levels of a $^1\Sigma^+$ molecule. The states $|J, m_J\rangle = |0, 0\rangle$ and $|1, 1\rangle$ are split by an energy $2 h B_v$ where $B_v$ is the rotational constant \cred{for states with vibrational quantum number $v$}, usually a few GHz. The MW field is blue-detuned with respect to the resonance by an amount $\Delta$. (b) Classically, a $\sigma^+$ field with circular polarization sets the molecular dipole spinning at the frequency of the MW field. (c) Potential energy curves of a pair of MW-dressed NaCs molecules approaching each other in the $s$-wave channel for $\Omega / (2 \pi) = 4$\,MHz and $\Delta / (2 \pi) = 6$\,MHz. The adiabatic potentials for $|{++}\rangle$ (blue solid line), $|{+0}\rangle$ (grey solid line), and $|{+-}\rangle$ (red solid line) are shown, which are the collisional channels based on the dressed states $|+\rangle$, $|-\rangle$ and the spectator states $|0\rangle$ of the individual molecules [for spectroscopy of the dressed states see \textcite{Zhang2024,Gu2025}]. Molecules are either (1) reflected by the repulsive potential, (2) lost to non-shielded states, or (3) reach short range. The inset shows the shallow potential well in the intermolecular potential that can support bound states. Adapted from \textcite{Bigagli2023}.}
    \label{fig:ms}
\end{figure}

As shown in Fig.~\ref{fig:ms}, such dressing through a MW field $\vector{E}(t)$ mixes rotational states to induce the dipole moment $\vector{d}_\text{ind}(t) = \bar{d}\left[\hat{\vector{e}}_+(t)\cos{\xi}+\hat{\vector{e}}_-(t)\sin{\xi}\right]$, where $\hat{\vector{e}}_\pm(t)$ are basis vectors rotating with the field's frequency that describe plus and minus circularly polarized fields. For example, $\xi = (0, \pi/4)$ correspond to a ($\sigma^+$, $\pi$) polarized MW. The resulting time-averaged interaction can be expressed as
\begin{equation}
    V_\text{dd}^{\xi}(r,\theta,\varphi) = \frac{\bar{d}^2}{4\pi\epsilon_0}\frac{3\cos^2{\theta}-1+3\sin{(2\xi)}\sin^2{\theta}\cos{(2\varphi)}}{r^3},
\label{eq:VddEliptical}
\end{equation}
using $\bar{d}=d_0/\sqrt{12\left[1+(\Delta/\Omega)^2\right]}$ \cite{Chen2023}, 
\cred{with $\Delta = \omega_\text{MW} - \omega_0$, where $\omega_\text{MW}$ is the MW frequency, $\omega_0$ is the rotational transition frequency}, and $\Omega$ is the Rabi frequency characterizing the coupling strength. 

For a linearly polarized MW, we can define $\vector{E}(t) = E(t) \, \hat{\vector{e}}_z$ such that $\hat{\vector{e}}_z$ sets the quantization axis, in analogy to the static-field case above. Then, the DDI can be expressed by Eq.~\eqref{eq:Vdd} using an effective dipole moment, which is $d_\text{eff} = d_0/\sqrt{6[1+(\Delta/\Omega)^2]}$ when the MW couples $\ket{0, 0}$ and $\ket{1, 0}$ with $\Omega = (d_0/\sqrt{3})E/\hbar$ \cite{Karman2022}. 

For a circular-polarized MW field, we can define the wave vector as the quantization axis. This procedure allows us to express the DDI as $-V_\text{dd}(\vector{r})$ using Eq.~\eqref{eq:Vdd} with $d_\text{eff} = d_0/\sqrt{12[1+(\Delta/\Omega)^2]}$ for the $\ket{0, 0}$-to-$\ket{1, \pm 1}$ transition \cite{Buechler2007,Micheli2007,Karman2022}. Due to the inverted sign of the DDI, this scenario is denoted as an ``anti-dipolar'' interaction. Specifically, in the anti-dipolar case, two circularly polarized dipoles experience a repulsive (attractive) DDI when approaching out of plane (in plane) of the rotating dipole, whereas in the regular dipolar case, two linearly polarized dipoles experience a repulsive (attractive) DDI under head-to-tail (side-by-side) configurations. 

For arbitrary polarizations, Eq.~\eqref{eq:VddEliptical} can also be rearranged as a linear combination
\begin{equation}
    V_\text{dd}^{\xi}(r,\theta,\varphi) = \left[1-\sin{(2\xi)}\right]V_\text{dd}^- + \sin{(2\xi)}V_\text{dd}^+
\label{eq:VddElipticalShort}
\end{equation}
of an anti-dipolar interaction $V_\text{dd}^- \propto -(1-3\cos^2{\theta})$ with $d_\text{eff}^-=\bar{d}$ pointing along the MW wave vector and an orthogonal normal dipolar interaction $V_\text{dd}^+ \propto (1-3\sin^2{\theta}\sin^2{\varphi})$ with $d_\text{eff}^+=\sqrt{2}\bar{d}$. At the magic MW ellipticity $\xi_\text{m}=\sin^{-1}{(1/3)}/2\approx 10^\circ$, we find antisymmetric interactions $V_\text{dd}^{\xi}(r,0,\varphi) = -V_\text{dd}^{\xi}(r,\pi/2,\pi/2)$ along two axes and vanishing interactions $V_\text{dd}^{\xi}(r,\pi/2,0) = 0$ along the third axis. 

We briefly note that anti-dipolar interactions and (limited) tuning of the DDI can also be achieved for magnetic atoms by a continuous rotation of the magnetic field \cite{Giovanazzi2002,Tang2018tuning,Baillie2020,Halder2022}.

\subsection{Intermediate-range barrier and shielding}
\label{sub:sec:shielding}

\cred{A central challenge in experiments with ultracold molecules is} that they undergo lossy collisions when they reach short range. This loss can be the result of an exothermic chemical reaction channel, such as for KRb \cite{Ospelkaus2010,Ni2010, Quemener2012,Hu2019} or CaF \cite{Cheuk2020,Sardar2023}. However, even molecules that are chemically stable with respect to two-body collisions \cite{Zuchowski2010,Smialkowski2021,Tomza2021,Ladjimi2024} can form sticky complexes at short range, which subsequently are subject to loss processes \cite{Mayle2013,Christianen2019,Liu2020,Gregory2020,Gersema2021,Bause2021,Bause2023}.
Harnessing the sizable long-range interaction of dipolar molecules opens a path to engineering a repulsive interparticle interaction that prevents colliding molecules from reaching short range. 
\cred{However, for the bare dipole–dipole interaction there is always at least one direction} along which the anisotropic long-range interaction is attractive. Hence, additional potential barriers have to be erected to prevent the collapse of molecular ensembles.
In the following, we present several possible schemes to do so, in order to shield molecules against collisional losses.

\paragraph{Confinement dc shielding} A first scheme is confinement shielding, which takes advantage of the anisotropy of the DDI. When a static electric field $\vector{E} = E \, \hat{\vector{e}}_z$ is used to induce a dipole moment in ground-state molecules, a strong trapping potential in the $\hat{\vector{e}}_z$ direction (e.g., a one-dimensional optical lattice) can be used to effectively confine the molecules to a plane orthogonal to the electric field axis \cite{Buechler2007,Micheli2007}. We assume that the angular trap frequency $\omega_z$ of the confinement is sufficiently strong and the temperature sufficiently low to limit the molecules to the motional ground state of this confinement. The spread of the molecular wavefunction in the confinement direction is then characterized by the harmonic oscillator length $a_\text{ho} = \sqrt{ \hbar / \mu \omega_z}$.
For $a_\text{d} \gg a_\text{ho}$ the oriented molecules mainly repel each other side by side at distances larger than $a_\text{ho}$, which renders the collisions effectively two-dimensional \cite{Ticknor2009,Quemener2010}. The repulsive DDI shields the molecules from reaching the lossy short-range regime \cite{Ticknor2010,Micheli2010,Quemener2011,Miranda2011,Julienne2011,Frisch2015,Quemener2015}.
For $a_\text{d} < a_\text{ho}$ the confinement alone is not sufficient to prevent attractive head-to-tail collisions when the molecules approach each other and the collision dynamics becomes more three-dimensional. \cred{The exception is in the case of indistinguishable fermions}\cred{, where the symmetrization principle of the wavefunction in confined geometries~\cite{Miranda2011,Quemener2011} suppresses the head-to-tail component of the $p$-wave centrifugal barrier, leaving only the repulsive side-by-side collisions and hence supporting the shielding.}
The latter setting has been used successfully by \textcite{Valtolina2020} to create a degenerate Fermi gas of dipolar ${}^{40}$K${}^{87}$Rb molecules.

As, however, confinement shielding is generally restricted to low-dimensional systems, it is useful to find shielding schemes that work in three dimensions. 

\begin{figure*}[tb]
    \centering
    \includegraphics[width=0.98\textwidth]{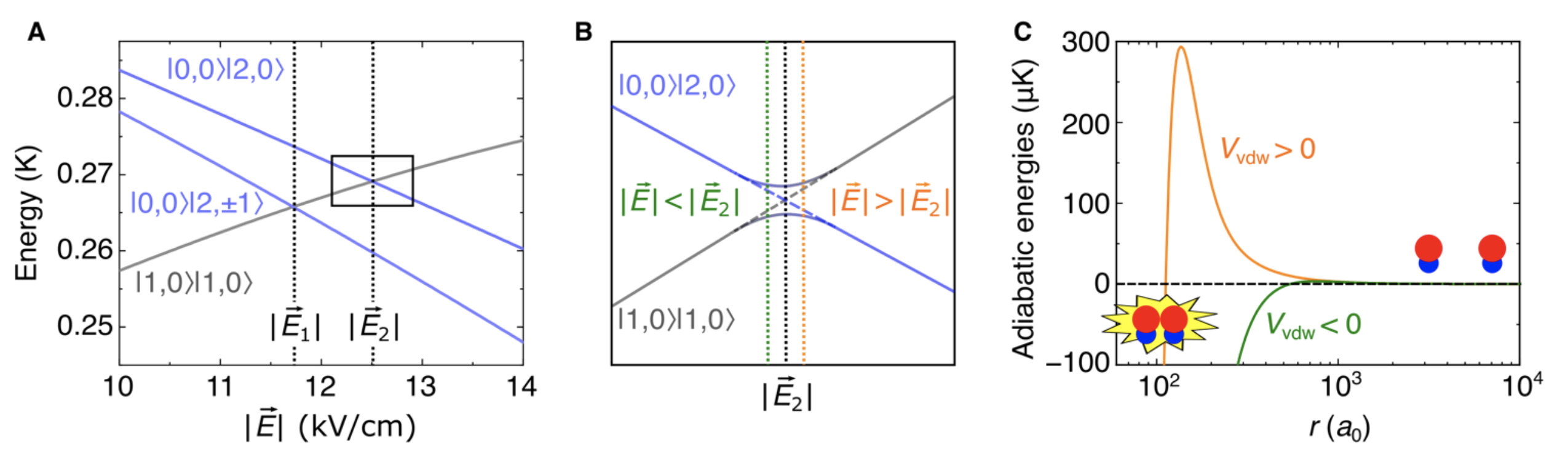}
\caption{Resonant static field shielding. (a) Collision channels as a function of electric field. Shielding can be realized when channels cross. (b) \cred{Zoomed-in view of such a crossing at a field} $\vector{E}_2$. Dipolar interactions lead to an avoided crossing between the channels. This avoided crossing leads to an effective van der Waals interaction. (c) Resulting adiabatic energies for two electric fields below (above) $|\vector{E}_2|$. For $|\vector{E}|>|\vector{E}_2|$ a significant barrier is formed that prevents the molecules from reaching short range, thus leading to shielding. Taken from \textcite{Matsuda2020} \cred{and shown here for KRb molecules.}}
    \label{fig:shieldingprinciple}
\end{figure*}

\paragraph{Resonant dc shielding} A second scheme that achieves this goal is resonant dc-field shielding. This technique takes advantage of a 
F{\"o}rster resonance between two scattering states, that are described by the combined rotational states 
$\ket{1} \equiv \ket{\tilde{1}, 0}\ket{\tilde{1},0}$ 
and $\ket{2} \equiv \left(\ket{\tilde{0},0}\ket{\tilde{2},0}+\ket{\tilde{2},0}\ket{\tilde{0},0}\right)/\sqrt{2}$
\cite{Avdeenkov2006,Wang2015,Quemener2016,GonzalesMartinez2017,Matsuda2020,Li2021,Lassabliere2022,Mukherjee2023b,Mukherjee2024,Mukherjee2025}
of energy $\varepsilon_{\ket{1}}$ and $\varepsilon_{\ket{2}}$, as illustrated in Fig.~\ref{fig:shieldingprinciple}. 
The notations $\ket{1}$ and $\ket{2}$ for the scattering states have been used for simplicity. 
These two states are degenerate (i.e., $\varepsilon_{\ket{1}} = \varepsilon_{\ket{2}}$) around an electric field of $E^* \approx 3.25\, B_\text{rot}/d_0$ \cite{GonzalesMartinez2017}.

To understand the shielding mechanism at a glance, one can approach the problem as a two-level system~\cite{Lassabliere2022,Li2021}. In this case, it can be shown that, to a good approximation, the DDI for the two scattering states around the field $E^*$ can be described by
\begin{multline}\label{eq:shielding1}
    V^{\pm}_\text{dc}(\vector{r}) \approx \frac{1}{2}\left[ V_\text{dd}(\vector{r},d_{11}^2+d_{00}d_{22}+d_{02}^2) +\varepsilon_{\ket{1}} + \varepsilon_{\ket{2}} \right] \\
    \pm \frac{1}{2} \Bigl\{\left[V_\text{dd}(\vector{r},d_{11}^2-d_{00}d_{22}-d_{02}^2) + \varepsilon_{\ket{1}} - \varepsilon_{\ket{2}} \right]^2 \\
    + 8 \left[ V_\text{dd}(\vector{r},d_{10}d_{12})\right]^2 \Bigr\}^{1/2},
\end{multline}
where $V_\text{dd}(\vector{r},D)$ corresponds to $V_\text{dd}(\vector{r})$ with $C_\text{dd} = D / \epsilon_0$.
One can define the transition dipole moment between states 
$\ket{\tilde{N}, m_N}$ and $\ket{\tilde{N}', m_N'}$ as 
$d_{NN'}^{{m_N} {m_N}'} = \bra{\tilde{N}, m_N} \vector{d}_0 \cdot \vector{e_z} \ket{\tilde{N}', m_N'}$, which is a generalization of the induced dipole moment. 
The quantities $d_{NN'}$ used here are the transition dipole moments for projections
 $m_N = m_N'=0$.
The terms $V^{\pm}_\text{dc}$ are reminiscent of a Born--Oppenheimer-like picture \cite{Micheli2007,Lassabliere2022}
where the two molecules collide on a unique potential energy surface (PES).
The $V_\text{dc}^+$ ($V_\text{dc}^-$) PES is mostly repulsive (attractive).
When $E > E^*$, the scattering state $\ket{1}$ is above $\ket{2}$ in energy
and adiabatically follows the $V_\text{dc}^+$ PES to a good approximation. 
When averaging $V_\text{dc}^+$ over the lowest partial waves $\ket{L, m_L}$,
the overall effective radial potential becomes repulsive \cite{Wang2015,Lassabliere2022}, providing a powerful shielding barrier, as shown in Fig.~\ref{fig:shieldingprinciple}(c).

\textcite{Mukherjee2025} provided an effective interaction potential in second-order perturbation theory, that has the form
\begin{equation}\label{eq:Vdcperp}
    V^{+}_\text{dc,pert}(r,\theta) = \frac{C_3 (\theta)}{r^3}+\frac{C_6(\theta)}{r^6},
\end{equation}
with a repulsion dominant at short-range, and DDIs at long-range. Here, and throughout, the coefficient $C_6$ is not to be confused with the usual van der Waals terms, which mainly originate from couplings to higher electronic states or rovibrational states. The authors fitted essential parameters such as $d_{NN'}$ and $\varepsilon_{\ket{1}} - \varepsilon_{\ket{2}}$ as functions of $d_0 E/B_\text{rot}$ with low-order polynomials to provide universal parameters that describe $V^{+}_\text{dc,pert}(r,\theta)$  [see Tab.~I of \textcite{Mukherjee2025}]. Moreover, they also provided a more precise effective potential
\begin{equation}
    V^{+}_\text{dc,fc}(r,\theta) = \frac{1}{2}(M_{11} + M_{22}) + \frac{1}{2}\sqrt{(M_{11} - M_{22})^2 + 4M_{12}^2},
\end{equation}
using Van Vleck perturbation theory to build a fully coupled (fc) model. The matrix elements $M_{ij}(r,\theta)$, where $i,j$ denote the states $\ket{1}$ or $\ket{2}$, include the elements that arise from the Van Vleck transformation, which can once again be expanded by universal polynomial parameters [see Tab.~II of \textcite{Mukherjee2025}].

Resonant dc shielding is typically efficient in the electric field range of $3.25 B_\text{rot}/d_0 \leq E \leq 3.8 B_\text{rot}/d_0$. This relatively narrow window reflects a key limitation of the technique, namely the restricted tunability of the interaction strength. Shielding is most effective around $E = E_\text{s} \approx 3.4 B_\text{rot}/d_0$ \cite{GonzalesMartinez2017} where a compromise is struck between maximizing the shielding barrier to avoid losses at short range and separating the states $\ket{1}$ or $\ket{2}$ to minimize state-changing collisions \cite{Lassabliere2022,Mukherjee2023b,Mukherjee2024}. In general, it works best for molecules with a high ratio of rotational constant to dipolar energy $\tilde{B}_\text{rot} = B_\text{rot} / \tilde{E}_\text{d}$ \cite{GonzalesMartinez2017}. While $\tilde{B}_\text{rot}$ is fairly small for KRb molecules, as shown in Tab.~\ref{tab:mwdressing}, \textcite{Matsuda2020,Li2021} managed to apply resonant dc shielding to fermionic ${}^{40}$K${}^{87}$Rb molecules in 2D and 3D systems. \cred{To extend beyond shielding of a single internal state, it} was suggested by \textcite{Mukherjee2025b,Mukherjee2025c} that resonant dc shielding is ideal to realize systems with SU$(N)$ symmetry by making use of the numerous hyperfine states in molecules.

\begin{table*}[]
\centering
\begin{ThreePartTable}
\caption{Field parameters for shielding of relevant bosonic molecular species in their electronic and vibrational ground state. The reduced rotational constant $\tilde{B}=B_\text{rot}/\tilde{E}_\text{d}$ is a measure for the effectiveness of the static electric field shielding. At the resonance field $E^*$ the molecules in $\ket{\tilde{1}, 0}$ have the induced dipole moment $d^*$ and the dipolar length $a_{\text{dd}}^*$. For microwave shielding we consider double MW shielding with fixed $\Omega_\sigma$, $\Omega_\pi$, and $\Delta_\sigma$, while $\Delta_\pi^\text{min}$ and $\Delta_\pi^\text{max}$ mark the limits of a 'stability window', defined as region without field-linked state. $a_\text{dd}^\text{min}$ and $a_\text{dd}^\text{max}$ are the corresponding dipolar lengths at these limits, where negative values denote anti-dipolar interactions. If no limits of $\Delta_\pi$ are stated, there is no field-linked state and $a_\text{dd}^\text{min}$ and $a_\text{dd}^\text{max}$ are given for $\Delta_\pi = 0$ and for pure circular-polarized MW shielding, respectively. For comparison, $a_\text{dd} = 130\,a_0$ for magnetic Dy atoms. The underlying molecular constants $d_0$ and $B_\text{rot}$ are listed in the appendix in Tab.~\ref{tab:const}. \cred{Detailed calculations of two-body loss rates for several molecular species under double-microwave-shielding conditions have been presented by~\textcite{Karman2025}, identifying the general trend that heavier and more strongly dipolar molecules tend to exhibit lower loss rates. Experimental rates have been measured for NaCs, NaRb and (fermionic) NaK~\cite{Yuan2025,Shi2025DAMOP,Schindewolf2022}.}}
\setlength{\tabcolsep}{5pt}
\begin{tabular}{l r@{}l r@{}l c c r@{}l r@{}l c c}
\hline
\hline
 &  &  &  &  &  &  & \multicolumn{4}{c}{Microwave shielding} &  &  \\
 & \multicolumn{4}{c}{Static-field shielding} &  &  & \multicolumn{4}{c}{$\Omega_\pi=\Omega_\sigma=10 \times 2\pi$\,MHz} &  &  \\
 &  &  &  &  &  &  & \multicolumn{4}{c}{$\Delta_\sigma=0$\,MHz} &  &  \\
\cline{2-5}\cline{8-11}\rule{0pt}{10pt}
Molecule & \multicolumn{2}{c}{$\tilde{B}$} & \multicolumn{2}{c}{$E^*$} & $d^*$ & $a_{\text{dd}}^*$ & \multicolumn{2}{c}{$\Delta_\pi^\text{min}/2\pi$} & \multicolumn{2}{c}{$\Delta_\pi^\text{max}/2\pi$} & $a_\text{dd}^\text{min}$ & $a_\text{dd}^\text{max}$ \\
  & \multicolumn{2}{c}{} & \multicolumn{2}{c}{(kV/cm)} & (D) & ($a_0$)  & \multicolumn{2}{c}{(MHz)} & \multicolumn{2}{c}{(MHz)} & ($a_0$) & ($a_0$)\\
\hline\rule{0pt}{10pt}%
${}^7$Li${}^{23}$Na      &  2&.$8\times10^5$    & \ 160&.50  & 0.06 &    11 & \ \ \ \  &    & \ \ \   &   &    36 &    -71 \\
${}^6$Li${}^{40}$K       &  2&.$5\times10^9$    & \  16&.60  & 0.48 &   979 & \ \ \ \  &    & \ \ \   &   &  3130 &  -6200 \\
${}^7$Li${}^{85}$Rb      &  2&.$8\times10^{10}$ & \  10&.49  & 0.56 &  2686 & \ \ \ \ 0&.11 & \ \ \  9&.9 &  8170 & -10800 \\
${}^7$Li${}^{133}$Cs     &  3&.$1\times10^{11}$ & \   6&.59  & 0.77 &  7803 & \ \ \ \ 1&.4  & \ \ \  5&.2 & 10700 & -15400 \\
${}^{23}$Na${}^{39}$K    &  8&.$4\times10^8$    & \   6&.72  & 0.38 &   855 & \ \ \ \  &    & \ \ \   &   &  2730 &  -5400 \\
${}^{23}$Na${}^{87}$Rb   &  6&.$4\times10^9$    & \   4&.22  & 0.45 &  2075 & \ \ \ \ 0&.19 & \ \ \ 11&   &  6100 &  -8950 \\
${}^{23}$Na${}^{133}$Cs  &  6&.$2\times10^{10}$ & \   2&.46  & 0.64 &  5950 & \ \ \ \ 1&.3  & \ \ \  5&.5 &  9300 & -13300 \\
${}^{41}$K${}^{87}$Rb    &  5&.$5\times10^6$    & \  12&.32  & 0.08 &    78 & \ \ \ \  &    & \ \ \   &   &   248 &   -492 \\
${}^{39}$K${}^{133}$Cs   &  1&.$2\times10^9$    & \   3&.17  & 0.26 &  1096 & \ \ \ \ 0&    & \ \ \ 20&   &  3500 &  -6000 \\
${}^{87}$Rb${}^{133}$Cs  &  2&.$6\times10^8$    & \   2&.58  & 0.17 &   608 & \ \ \ \  &    & \ \ \   &   &  1940 &  -3850 \\
${}^{39}$K${}^{107}$Ag   &  7&.$1\times10^{11}$ & \   1&.52  & 1.19 & 19428 & \ \ \ \ 1&.9  & \ \ \  4&.1 & 16400 & -23500 \\
${}^{87}$Rb${}^{107}$Ag  &  1&.$2\times10^{12}$ & \   0&.78  & 1.26 & 29133 & \ \ \ \ 2&.1  & \ \ \  3&.8 & 18500 & -26100 \\
${}^{133}$Cs${}^{107}$Ag &  2&.$2\times10^{12}$ & \   0&.53  & 1.37 & 42025 & \ \ \ \ 2&.3  & \ \ \  3&.5 & 19700 & -27900 \\
${}^{87}$Rb${}^{88}$Sr   &  3&.$3\times10^8$    & \   2&.36  & 0.21 &   725 & \ \ \ \  &    & \ \ \   &   &  2300 &  -4600 \\
${}^{6}$Li${}^{53}$Cr    &  7&.$5\times10^9$    & \  27&.43  & 0.46 &  1184 & \ \ \ \  &    & \ \ \   &   &  3710 &  -7400 \\
${}^{40}$Ca${}^{19}$F    &  4&.$1\times10^9$    & \  21&.59  & 0.43 &  1024 & \ \ \ \  &    & \ \ \   &   &  3270 &  -6500 \\
${}^{88}$Sr${}^{19}$F    &  3&.$0\times10^{10}$ & \  13&.83  & 0.49 &  2409 & \ \ \ \ 0&.34 & \ \ \  9&.8 &  6600 &  -9600 \\
${}^{138}$Ba${}^{19}$F   &  5&.$6\times10^{10}$ & \  13&.18  & 0.44 &  2907 & \ \ \ \ 0&.72 & \ \ \  7&.6 &  6500 &  -9400 \\
${}^{174}$Yb${}^{19}$F   &  2&.$7\times10^{11}$ & \  11&.94  & 0.55 &  5438 & \ \ \ \ 1&.3  & \ \ \  5&.5 &  8300 & -11900 \\
${}^{27}$Al${}^{19}$F    &  1&.$9\times10^8$    & \  70&.26  & 0.21 &   195 & \ \ \ \  &    & \ \ \   &   &   620 &  -1230 \\
${}^{89}$Y${}^{16}$O     &  1&.$2\times10^{11}$ & \  16&.60  & 0.63 &  3958 & \ \ \ \ 0&.82 & \ \ \  7&.2 &  8380 & -12100 \\
\hline
\hline
\end{tabular}
\label{tab:mwdressing}
\end{ThreePartTable}
\end{table*}

\paragraph{Microwave and optical ac shielding} Using avoided crossings at an intermediate distance between attractive and repulsive scattering-state curves is not specific to resonant dc shielding, but instead reflects a more general principle underlying various three-dimensional shielding techniques. Another example of such shielding is to use an ac electric field instead of a dc field. This third scheme takes advantage of the flexibility that an electromagnetic wave provides when dressing a molecule. The detuning $\Delta$ of the electromagnetic wave can vary depending on the energy of the corresponding photon, and its Rabi frequency $\Omega$ related to the strength of the corresponding ac field, and its polarization.

Many different methods along these lines have been theoretically proposed and investigated. These studies are typically based on either optical shielding, where a laser field excites electronic states, or microwave shielding, where a MW field dresses rotational states within the electronic ground state.

Although optical shielding~\cite{Suominen1995,Napolitano1997} has been explored extensively, it has so far proven to be ineffective, primarily due to photon scattering and unwanted heating in a one-photon scheme~\cite{Xie2020}, and because second-order couplings are too weak to generate a sufficiently repulsive shielding curve in a two-photon scheme~\cite{Karam2023,Karam2024,Karam2025}.

In contrast, microwave shielding has emerged as a more promising approach. By coupling rotational states within the electronic ground state, it enables strong and tunable interactions while minimizing losses. Following a variety of theoretical proposals \cite{Buechler2007,Micheli2007,Gorshkov2008,Cooper2009,Huang2012,Lassabliere2018,Karman2018,Karman2019,Karman2020}, it has been observed and implemented in a variety of experiments \cite{Anderegg2021,Bigagli2023,Lin2023} and has been a crucial tool for the realization of stable ultracold molecular gases~\cite{Schindewolf2022, Bigagli2024,Shi2025DAMOP}.

\begin{figure*}
    \centering
    \includegraphics[width=\textwidth]{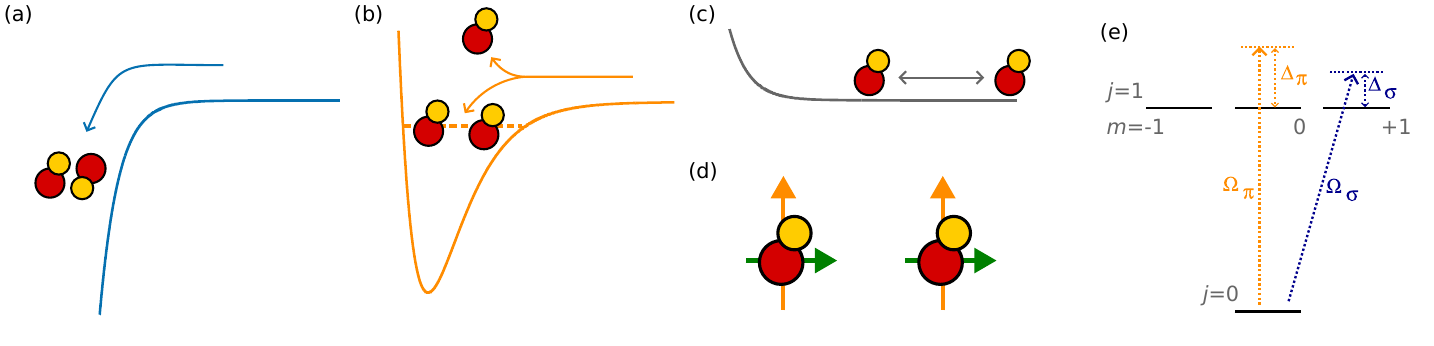}
    \caption{Molecular collision dynamics and double MW shielding. (a) No shielding, where molecules interact by attractive van der Waals interactions and undergo loss at short range. (b) Single-field MW shielding, where molecules are shielded from two-body collisional loss but can undergo three-body recombination into field-linked bound states. (c) Double MW shielding, where the dipolar interaction between shielded molecules can be tuned or even compensated, resulting in a repulsive potential that does not support any bound states. (d) Illustration of long-range interactions between double MW shielded molecules, which can be thought of as the sum of independent interactions between a lab-frame dipoles in the $z$ direction (orange) and, for finite ellipticity, lab-frame dipoles in the $y$ direction (green). Here, the $z$ direction is the propagation direction of the elliptically polarized microwave and the polarization direction of the linearly polarized microwave. The two dipolar interactions can be controlled using the MW-field intensities,  frequencies and ellipticities. Taken from \textcite{Karman2025}.}
    \label{fig:dms}
\end{figure*}

To establish microwave shielding, a circularly polarized (i.e., $\xi = 0$, $\pi$ for $\sigma^+$, $\sigma^-$ polarization) MW field is used, which is blue detuned from the $| 0, 0 \rangle \rightarrow | \xi_+ \rangle = \cos{\xi} | 1, 1 \rangle + \sin{\xi} | 1, -1 \rangle$ transition and transfers the molecules into a dressed state $| + \rangle = u | 0, 0 \rangle + v | \xi_+ \rangle$ [cf. Fig.~\ref{fig:ms}(a)]. Here, $u = \sqrt{(1+\Delta/\Omega_\text{eff})/2}$, $v = \sqrt{(1-\Delta/\Omega_\text{eff})/2}$, and the effective Rabi frequency $\Omega_\text{eff} = \sqrt{\Omega^2 + \Delta^2}$ \cite{Deng2023}. The blue detuning ensures that the molecules collide in $|{++}\rangle$, the energetically highest state of a manifold of collisional dressed states [see Fig.~\ref{fig:ms}(c)]. At long range Eq.~\eqref{eq:VddEliptical} describes the interaction in this collisional channel. When the molecules approach along the attractive directions of $V_\text{dd}^\xi(r,\theta,\phi)$ they start coupling to repulsively interacting lower-lying collisional channels, which feature molecules in the other dressed state $|-\rangle$ or in the so-called spectator states $|0\rangle$, \cred{which are dark states for single-molecule MW dressing. The} remaining inelastic processes originate either from tunneling into the short range or, more likely, from predissociation into lower-lying dressed states, as illustrated in Fig.~\ref{fig:ms}(c). For moderate ellipticity ($\xi < \pi/6$) the interaction potential including the repulsive barrier has been approximated \cite{Deng2023,Xu2025} using perturbation theory as
\begin{multline}
    V_\text{MW}(r,\theta,\varphi) \approx V_\text{dd}^\xi(r,\theta,\varphi)\\
    + \frac{C_6}{r^6}\sin^2{\theta}\Bigl\{1 - \sin^2(2\xi)\cos^2(2\varphi)\\
    + \left[ 1 - \sin(2\xi)\cos(2\varphi) \right]^2 \cos^2{\theta} \Bigr\},
    \label{eq:Vpp1MW}
\end{multline}
with $C_6 = d_0^4/\left\{128\pi^2\epsilon_0^2\hbar\Omega[1+(\Delta/\Omega)^2]^{3/2}\right\}$. The first term can also again be rewritten using Eq.~\ref{eq:VddEliptical} in terms of a dipolar $C_3/r^3$ term multiplied with an angular dependence as in Eq.~\ref{eq:genericpotential}. In contrast to resonant dc shielding, where the interaction parameters can only be tuned through $E$, the MW-shielding interaction potential is highly tunable through the MW-field parameters $\Omega$, $\Delta$, and $\xi$.
\textcite{Dutta2025} demonstrated that the elastic and inelastic scattering can universally be described for various molecular species after normalizing the involved energies by $\tilde{E}_\text{d}$ and the length scales by $R_3 = \tilde{a}_\text{d}$, as long as $\Omega, \Delta \ll B_\text{rot}/\hbar$. Consequently, molecules with large $d_0$ and large mass are, in principle, best suited for MW shielding.

So far, we have considered molecules in an electronic singlet ground state ${}^1\Sigma$, such as bialkalis. In this case, the coupling between the states that form the shielding potential typically dominates over the hyperfine structure splittings. However, for molecules that feature an unpaired electron, e.g., laser-coolable molecules with a ${}^2\Sigma$ ground state, the fine and hyperfine structure in the ground state can have a comparable or dominating energy splitting. While the effects of the unpaired spins are minor for resonant dc shielding \cite{Quemener2016,GonzalesMartinez2017,Mukherjee2023b} they are more relevant for MW shielding. Applying a modest magnetic field can help suppress hyperfine-inelastic collisions \cite{Lassabliere2018,Karman2018,Karman2019}, however, it can also break the singlet coupling between electronic and nuclear spin~\cite{Anderegg2021}\cred{, which can alter the hyperfine structure relevant for the shielding scheme.}
Selecting the spin-stretched transition with the largest energy can help to maintain a blue detuning with respect to other transitions~\cite{Karman2018}. \cred{Overall, these considerations highlight that the detailed internal spin structure of the molecules must be taken into account when designing microwave-shielding schemes.}

\paragraph{Double microwave ac shielding} An even more tunable extension of the regular MW shielding is double microwave shielding \cite{Deng2025,Karman2025}. This technique features an extreme suppression of both, inelastic two-body collisions and three-body recombination (see Sec.~\ref{sub:sec:three-body}), as well as a wide tuning range of dipolar interactions~\cite{Yuan2025}, which has been essential for the first realizations of BECs of dipolar molecules \cite{Bigagli2024,Shi2025DAMOP}. Here, a second blue-detuned MW field with a frequency that typically differs by at least $2\pi \times 2$\,MHz from the first field is used. This second field is $\pi$ polarized (i.e., orthogonal to the circular or elliptical field) and, hence, addresses the $| 0, 0 \rangle \rightarrow | 1, 0 \rangle$ transition, as illustrated in Fig.~\ref{fig:dms}(e). In essence, it induces an effective regular dipolar interaction, which in itself does not provide a 3D shielding potential but can be used to cancel out the anti-dipolar long-range interaction induced by the circularly polarized shielding MW. The combination of the two fields maintains a shielding barrier at intermediate range while suppressing the long-range dipolar interaction [see Fig.~\ref{fig:dms}(c)]. For $\xi = 0$ the coupling of the MW fields to the individual molecules is effectively a coupled 3-level system, where the eigenvectors can be expressed with the nontrivial Euler mixing angles $\alpha$, $\beta$, and $\gamma$. Together with the coefficients $w_0$, $w_1$, and $w_2$ from the appendix of \textcite{Deng2025}, an effective potential can be expressed using perturbation theory as
\begin{multline}
    V_\text{doubleMW}(r,\theta) \approx\\
    \frac{C_6}{r^6}\Big[w_2\sin^4{\theta} + w_1\sin^2{\theta}\cos^2{\theta}
    + w_0\left(3\cos^2{\theta}-1\right)^2\Big]\\
    + \frac{C_3}{r^3}\left(3\cos^2{\theta} - 1\right),
    \label{eq:Vpp2MW}
\end{multline}
with $C_6=d_0^4/(4\epsilon_0^2)$ and
\begin{equation}
    C_3 = \frac{d_0^2}{24\pi\epsilon_0}\left[3\cos{(2\beta)} - 1\right] \cos^2{\alpha}\sin^2{\alpha}.
\end{equation}
The angle $\alpha$ is a measure for the contribution of the excited rotational states to $| + \rangle$, while $\beta$ is a measure for the contribution of the individual excited states ($\beta = 0$ for only $| 1, \pm1 \rangle$ and $\beta = \pi/2$ for only $| 1, 0 \rangle$). Missing in this description are interactions from the exchange of photons between the two MW fields (in a Floquet picture). These additional couplings can play a significant role when the beat frequency between the MW frequencies matches the separation of the MW-dressed scattering channels and should be avoided~\cite{Karman2025}. Deviations from ideal $\sigma$ and $\pi$ polarization, which results in two orthogonal effective lab-frame dipole moments, as illustrated in Fig.~\ref{fig:dms}(d), were also covered by \textcite{Karman2025}. Alternative possibilities for tuning the potentials have been proposed, either by addressing the $\pi$-polarized MW $N=1 \rightarrow 2$ transition or by replacing it with a static electric field \cite{Micheli2007,Gorshkov2008,Schmidt2022,Xu2025}.

\subsection{Field-linked resonances and tunability}\label{sub:sec:FLresonances}

For magnetic atoms, Feshbach resonances~\cite{Kohler2006,Chin2010} provide a tool to tune the scattering phase independently of the long-range DDI by coupling to a short-range bound state. This gives rise to an effective short-range contact interaction, which, for ultracold bosons, is characterized by the $s$-wave scattering length $a_s$. The competition between the contact interaction and the DDI brings about the rich many-body physics that we discuss in \cred{Sec.~\ref{sec:many-body}}. Feshbach resonances have also been found for the special case of light NaLi molecules in their weakly dipolar triplet ground state \cite{Park2023b}, but are not widely available for polar molecules. 

\cred{Even in the absence of accessible Feshbach resonances, the} scattering potentials of the shielding methods discussed above can naturally be deformed through the external-field parameters (e.g., $|\vector{E}|$ for resonant dc shielding and $\Omega$, $\Delta$, and $\xi$ for MW shielding). This translates into tunability of the scattering phase shift and scattering length.

\begin{figure}
    \centering
    \includegraphics[width=0.97\columnwidth]{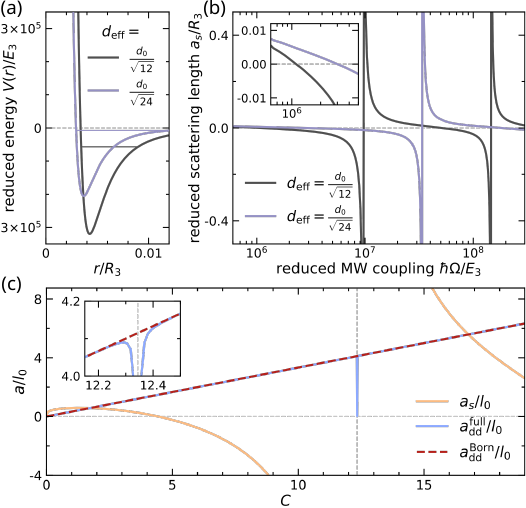}
    \caption{Field-linked states and tunability of interactions with single-MW shielding. (a) The adiabatic potential of the $s$-wave scattering channel as a function of the intermolecular distance $r$ for $\hbar\Omega/E_3 = 7 \times 10^7$ with a resonant MW field (black) and with a MW detuning $\Delta = \Omega$ (purple). The solid horizontal lines mark energies of field-linked states. Energies are normalized by $E_3 = \tilde{E}_\text{d}$ and lengths by $R_3 = \tilde{a}_\text{d}$. (b)~Universal tunability of the $s$-wave scattering length $a_s$ for $\Delta = 0$ (black) and $\Delta = \Omega$ (purple) at a collision energy of $1000\,E_3$. The data are taken from \textcite{Dutta2025}. (c)~Dependence of $a_s$ and the dipolar length $a_\text{dd}$  across the first field-linked resonance on the dimensionless interaction strength $C$ from Eq.~\eqref{eq:C} for the effective potential of Eq.~\eqref{eq:Vpp1MW} with $\xi=0$. The red dashed line and the solid blue line represent the Born approximation 
and the full calculation of $a_{\text{dd}}$, respectively\cred{, as discussed in Sec.~\ref{sub:sec:effV}}. The two curves show excellent agreement except within a small region near the resonance (see inset). Both lengths have been scaled by $l_0 = (C_6/C_3)^{1/3}$.}
    \label{fig:fieldlinkedtuningsingleMW}
\end{figure}

If the potential well that forms between the attractive long-range potential and the shielding barrier is sufficiently deep, it can support long-range weakly-bound states, so-called \textit{field-linked} states [see Fig.~\ref{fig:fieldlinkedtuningsingleMW}(a)]. When these states emerge, they give rise to \textit{field-linked} resonances, at which the scattering length diverges [see Fig.~\ref{fig:fieldlinkedtuningsingleMW}(b)]. In contrast to Feshbach resonances, no second potential is required and the potential well only exists in the presence of the external field.

Originally, field-linked resonances and states were predicted for molecules that feature Lambda doubling in the presence of static electric fields, such as ${^{16}}$OH \cite{Avdeenkov2002,Avdeenkov2003,Avdeenkov2004}, ${^{16}}$OD \cite{Avdeenkov2005}, and CaOH in a low-field-seeking state~\cite{Augustovicova2019}.
Later, field-linked states were also predicted in the scenario of MW shielding \cite{Cooper2009,Huang2012,Lassabliere2018} and of resonant 
dc shielding \cite{Mukherjee2024}.
   
Experimentally, the first field-linked states were observed with MW-shielded fermionic Na${^{40}}$K molecules by \textcite{Chen2023,Chen2024,Gu2025}. While the depletion in the interaction potential $V_\text{MW}$ of Eq.~\eqref{eq:Vpp1MW} was not deep enough to support bound states for circular polarization ($\xi = 0$), turning to elliptical polarization closed the avoided crossing between the dressed states along one direction enough to support up to two field-linked states \cred{\cite{Chen2023,Chen2024}}. For MW-shielded bosonic molecules field-linked states are even more accessible, as demonstrated by \textcite{Stevenson2024,Yuan2025,Zhang2025b}.

While for Feshbach resonances complex short-range physics determines the scattering properties, the size of the shielding barrier, which is typically on the order of $10^3\,a_0$, limits the scattering physics to intermediate and long interparticle distances in shielded molecules. 

On the one hand, this introduces challenges for the theoretical modeling of molecular gases when the size of the shielding core becomes comparable to the interparticle distance or dominates over the resulting scattering length, which will be discussed in more detail in Sect.~\ref{sec:many-body}. On the other hand, ignoring short-range physics leads to a universal description of the particle interaction that depends only on the characteristics of the external field and on a few properties of the molecules, such as mass, dipole moment, and rotational constant. For single-MW shielding, \textcite{Dutta2025} demonstrated that the binding energies and scattering properties are universal when energies are normalized by the dipolar energy $E_3 = \tilde{E}_\text{d}$ and length scales by $R_3 = \tilde{a}_\text{d}$, as illustrated in Fig.~\ref{fig:fieldlinkedtuningsingleMW}. This universality is also applicable to double MW shielding \textcite{Karman2025}. Based on the effective potential given in Eq.~\eqref{eq:Vpp1MW} \textcite{Ciardi2025} approximated for $\xi=0$ that the first field-linked resonance appears at
\begin{equation}\label{eq:boundstate}
    \frac{M\Omega^2}{48\pi\epsilon_0\hbar^2}\left(\frac{8\pi\epsilon_0\hbar d_0^4}{3(\Omega^2 + \Delta^2)^{5/2}}\right)^{1/3} \approx 12.35,
\end{equation}
as shown in Fig.~\ref{fig:fieldlinkedtuningsingleMW}(c). Universal expressions of the tunable binding energy and scattering length $a_s$ have been approximated by \textcite{Li2025}.
Note that coupling to lower-lying dressed states increases near a field-linked resonance, such that also inelastic collisions are enhanced. \textcite{Chen2023} used this signature to map out the first field-linked resonances.

\begin{figure}
    \centering
    \includegraphics[width=0.96\columnwidth]{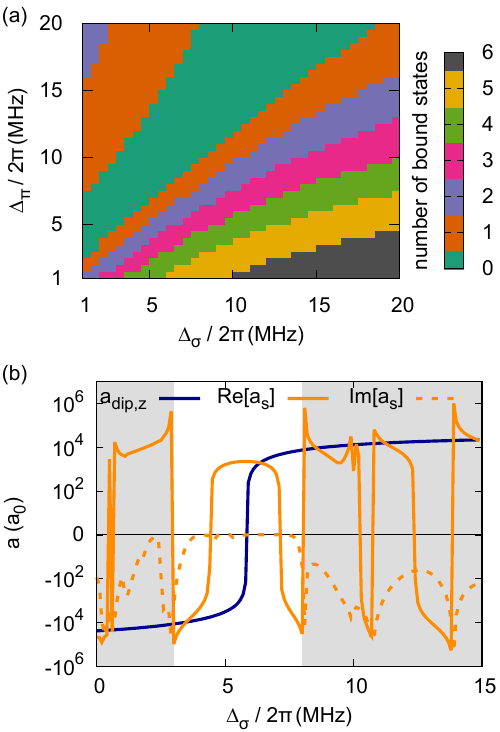}
    \caption{Controlling the field-linked bound states with double MW shielding. (a) Number of field-linked states for NaCs molecules controlled by the detunings of the $\pi$ and $\sigma$ MW fields at $\Omega_\sigma = \Omega_\pi = 10 \times 2\pi\,$MHz. (b) Scattering length and dipolar length, calculated as a function of detunings in the double MW shielding scheme for NaCs. Calculation is performed at 100~nK for $\Omega_\sigma = 10 \times 2\pi$~MHz, $\Omega_\pi = 10 \times 2\pi$~MHz and $\Delta_\pi = 10 \times 2\pi$~MHz. Results shown are for zero ellipticity. The data is adapted from \textcite{Karman2025}.} 
    \label{fig:tuningofinteractions}
\end{figure}

An example calculation for double MW shielding of NaCs is presented in Fig.~\ref{fig:tuningofinteractions}. The resulting scattering length notably shows several zero crossings, allowing access \cred{to} the full range of interactions, from attractive to non-interacting to repulsive, independent of $a_\text{dd}$.

\subsection{Three-body collisions}
\label{sub:sec:three-body}

While inelastic two-body collisions are enhanced near field-linked resonances, a more severe restriction of the stable parameter regime stems from three-body recombination: In parameter regimes that support field-linked states two molecules can bind to a field-linked state while a third molecule enables the exchange of binding and kinetic energy by providing momentum conservation \cite{Stevenson2024,Yuan2025} [see Fig.~\ref{fig:dms}(b)]. Suppressing field-linked states to achieve stable shielding limits the effectively available dipole moment of bosonic molecules to around $1$\,D. Table~\ref{tab:mwdressing} provides boundaries of the stable MW shielding regime for various dipolar molecules. For species with smaller dipole moments and masses, single-MW shielding is sufficient and the boundary $\Delta_\pi^\text{max}$ is omitted in Tab.~\ref{tab:mwdressing}. In contrast to this, species with larger dipole moments and masses require double MW shielding to suppress the field-linked states that start to emerge in this case, while simultaneously providing reasonable two-body shielding conditions. The stable parameter regime of NaCs is illustrated in Fig.~\ref{fig:tuningofinteractions}. For resonant dc shielding of molecules with very large dipole moments, the scarcity of tuning knobs makes it essentially impossible to suppress field-linked states while maintaining a decent ratio of elastic to inelastic two-body collisions, as calculated for NaCs, ${}^{39}$KAg, and CsAg by \textcite{Mukherjee2024}. They also showed that NaRb provides a rare exception where the first field-linked state is accessible in a certain field regime, which is embedded in a larger regime with favorable two-body shielding conditions.

If no field-linked states are supported, the role of three-body collisions remains an open question, warranting further study. For double MW-shielded NaCs molecules inelastic loss appears to be limited by (highly-suppressed) inelastic two-body collisions \cite{Yuan2025}. To what extent the shielding barrier holds for three-body collisions has only been discussed in simplified scenarios for single-MW shielding \cite{Huang2012} and resonant dc-field shielding \cite{Lassabliere2021}. 
For bosons, three-body collisions might enable coupling to dressed states that are antisymmetric under the exchange of two collision participants \cite{Gorshkov2008}. This would assist inelastic transitions to lower-lying dressed states. However, \textcite{Karman2025} illustrated that double MW shielding should suppress resonant transitions of this kind.

\subsection{From few-body to many-body bound states}
\label{sub:sec:few2many-body}
In analogy to magnetoassociation at atomic Feshbach resonances \cite{Kohler2006, Chin2010}, field-linked resonances have enabled coherent \textit{electroassociation} \cite{Quemener2023} of diatomic molecules into weakly bound states of tetramers, with transfer efficiency depending on the motional states and quantum degeneracy \cite{Chen2024, Deng2024}. These states are macroscopic in size with wavefunctions that can spread from $10^3\,a_0$ to $10^4\,a_0$ \cite{Chen2023}. The lifetime of the field-linked states is essentially limited by pre-dissociation into the continuum of the lower lying dressed scattering channels. So far, lifetimes of up to $8$\,ms have been realized by \textcite{Chen2024}, but lifetimes beyond $10$\,s should be feasible for molecules with larger dipole moments \cite{Deng2024} and with double MW shielding. Field-linked molecules might be a promising initial state for transfer \cite{Gacesa2021,Lepers2013} into tightly bound molecular states of tetramers \cite{Byrd2012,Christianen2019b,Yang2020FL,Sardar2023}, potentially, through shielded electronically excited states \cite{Mukherjee2025d}.

It is likely that larger field-linked states composed of more than two molecules also exist \cite{Wang2011,Huang2012,Shi2025b,Wang2025b}. Such few-body bound states might bridge the gap to strongly correlated many-body states, which we discuss in the next section. 

\section{Many-body physics and theory challenges}
\label{sec:many-body}
The long-range and anisotropic nature of dipolar interactions can give rise to a rich variety of exotic quantum phases in ultracold Bose gases. We begin by outlining the mean-field description of dipolar \cred{BECs}, which successfully captures the essential physics of magnetic quantum gases \cite{Lahaye2009,Boettcher2021,Chomaz2023}. We then address the limitations of this mean-field framework and its eventual breakdown as the dipolar interaction strength increases, driving the system into the strongly correlated regime. Finally, we discuss open questions and challenges in exploring this domain, along with first insights obtained from Quantum Monte Carlo simulations.

\subsection{Effective two-body interactions}
\label{sub:sec:effV}
In dilute and ultracold quantum gases, their interactions, such as, e.g., those discussed in the previous section, can be described by pseudopotentials \cite{weidemueller2003}.
This effective interaction is used in the mean-field theory discussed below, in which the microscopic potential is replaced by the two-body $T$-matrix at low energies, which provides an exact description of binary collisions, including bound-state resonances \cite{Dalfovo1999}.
For dipolar particles, the scattering amplitude of low-energy collisions can be accurately captured by the effective potential \cite{Marinescu1998,Yi2000,Yi2001}
\begin{align}\label{eq:PseudoPot}
V({\bf r}) = \frac{4\pi\hbar^2 a_s}{M}\delta({\bf r}) + V_3({\bf r}),
\end{align}
which combines the contact interaction characterized by the $s$-wave scattering length $a_s$ with the long-range dipole–dipole part of the interaction potential that scales as $\propto 1/r^3$ at large distances [see Eq.~\eqref{eq:genericpotential}]. The asymptotic form ($r\rightarrow\infty$) of all potentials presented in Sec.~\ref{sub:sec:shielding} can be expressed in this way [cf.\ Eqs.~\eqref{eq:shielding1}--\eqref{eq:Vpp2MW}]. 
The $s$-wave scattering length in Eq.~\eqref{eq:PseudoPot} contains contributions from both the short-range part (the respective $\propto C_6/r^6$ term) of the original interaction potential and the long-range dipole-dipole interactions (the respective $\propto C_3/r^3$ term). Contributions from higher order partial-wave channels are, within the first Born approximation, well captured by $V_3({\bf r})$, with small deviations occurring only in a narrow range around field-linked resonances [see Fig.~\ref{fig:fieldlinkedtuningsingleMW}(c)].  
The accuracy of this simple pseudopotential has first been demonstrated numerically for alkaline atoms with electric-field-induced dipole interactions \cite{Marinescu1998,Yi2000,Yi2001}, and later confirmed for magnetic dipolar atoms \cite{Oldziejewski2016} as well as microwave-dressed molecules \cite{Ciardi2025,Cardinale2025,Xu2025} for the type of MW-dressed interactions discussed in Sec.~\ref{sec:few-body}.

The strength of interactions can then be parametrized by the contact-interaction strength $g_s = 4\pi\hbar^2a_s/M$ and the dipole-dipole interaction strength $g_\text{dd} = 4\pi\hbar^2a_\text{dd}/M = C_\text{dd}/3\propto C_3$, such that the relative importance of dipole-dipole interactions is quantified by the parameter $\varepsilon_\text{dd} = g_\text{dd} / g_s=a_\text{dd}/a_s$. 
In the case of magnetic atoms, $\varepsilon_\text{dd}$ can be tuned by varying the scattering length $a_s$ via magnetic fields. For molecules, such a tuning of $a_s$ is also possible, as discussed in Secs.~\ref{sub:sec:FLresonances} and \ref{sub:sec:effV}, within the range constrained by the emergence of field-linked states and the corresponding losses [see Sec.~\ref{sub:sec:three-body}]. 
Similarly, the dipolar length $a_{\rm dd}$ can also be controlled directly via external electric fields and microwave fields.

Notably, for a given shielding scheme and field configuration, the distance dependence of the generated interaction potentials discussed in Sec.~\ref{sub:sec:shielding} can also be quantified by the single dimensionless parameter 
\begin{align}\label{eq:C}
C=\frac{MC_3^{4/3}}{\hbar^2C_6^{1/3}},
\end{align}
upon scaling with the characteristic length $\ell_0=(C_6/C_3)^{1/3}$ and the characteristic energy $\mathcal{E}_0=\hbar^2/(M\ell_0^2)$ \cite{Ciardi2025}. The two length scales $a_s/\ell_0$ and $a_{\rm dd}/\ell_0$ of the interaction are not independent parameters as both depend on $C$.

The pseudopotential of Eq.~\eqref{eq:PseudoPot} remains well applicable in the dilute-gas limit of small densities, in which the typical interparticle spacing greatly exceeds the characteristic length $\ell_0$ of the molecular potential, but it breaks down at higher densities as we will discuss in Sec.~\ref{sub:sec:qmc}. In this case, it needs to be replaced with the full potentials that describe, in particular, the intermediate-range size of typical shielding cores. Further approximations that capture the corresponding effects, e.g., in generalized mean-field descriptions, are a topic of ongoing research.

\subsection{Extended Gross--Pitaevskii equation}\label{sub:sec:eGPE}
\cred{A Bose–Einstein condensate (BEC) forms when a dilute gas of bosonic particles is cooled to sufficiently low temperatures such that a macroscopic fraction of the particles occupies a single quantum state \cite{Ketterle1999}. In this regime, the many-body system can be described by a macroscopic wavefunction, or order parameter, $\psi$, whose dynamics capture the collective behavior of the condensate. Dilute and weakly interacting dipolar BECs are well described by mean-field theory in terms of the extended Gross–Pitaevskii equation (eGPE)}
\cite{Wachtler2016,Wachtler2016b}
\begin{align}\label{eq:eGPE}
  i \hbar\, \partial_t \psi({\bf r}) =& \biggl(-\frac{\hbar^2 \nabla^2}{2M} + V_{\rm ext}({\bf r}) + \int V({\bf r}-{\bf r}^\prime)|\psi({\bf r}^\prime)|^2{\rm d}^3r^\prime\nonumber\\
  &+ \gamma_{\rm qf} |\psi({\bf r})|^3 - i \sum_{\alpha}\frac{\hbar L_\alpha}{2} |\psi({\bf r})|^{2\alpha-2} \biggr) \psi({\bf r}),
\end{align}
\cred{which governs the time evolution of this order parameter.} Within its domain of validity, the eGPE is system-agnostic and can be applied to both atoms and molecules. 

Here, ${V_{\rm ext}({\bf r}})$ denotes the confinement potential and typically takes the form of an anisotropic harmonic trap, and two-body interaction is given by Eq.~\eqref{eq:PseudoPot}.  The wave function $\psi$ is normalized to the particle number ${\mathcal N=\int \mathrm{d}^3r\, |\psi|^2}$.

Effects of quantum fluctuations can be included through the leading-order \cred{
Lee--Huang--Yang (LHY)} correction to the equation of state of a dipolar condensate \cite{Schutzhold2006,Lima2011,Lima2012} that describes interactions with excitations of the condensates \cite{Lee1957a,Lee1957b}. For a homogeneous system with constant condensate density $|\psi|^2$, the energy correction can be written as $\gamma_\mathrm{qf}|\psi|^3$\cred{, which in trapped systems is typically
applied locally within the local-density approximation (LDA). Here,} ${\gamma_\mathrm{qf} = (32/3\sqrt{\pi}) g_s a_s^{3/2} Q_5 (\varepsilon_\text{dd})}$ and ${Q_5(\varepsilon_\text{dd}) = \int_0^1 \mathrm{d}u\, (1-\varepsilon_\text{dd}+3\varepsilon_\text{dd}u^2)^{5/2}}$ \cite{Lima2011,Lima2012}. The function $Q_5$ is well approximated by its leading order Taylor expansion $Q_5(\varepsilon_\text{dd}) \sim 1+3\varepsilon_\text{dd}^2/2$ for $\varepsilon_\text{dd}\lesssim 1$ \cite{Bisset2016}. When $\varepsilon_\text{dd}$ exceeds unity, long-wavelength excitations become unstable such that $\gamma_\mathrm{qf}$ acquires an imaginary part. This issue is often circumvented by introducing infrared momentum cutoffs \cite{Wachtler2016,Bisset2016,He2024} or by simply discarding the imaginary part. In the strongly dipolar regime, the imaginary contribution to $\gamma_{\rm qf}$ can become significant and render such simple solutions questionable, as we will discuss in more detail in Sec.~\ref{sub:sec:limitsofmeanfield}.

Particle losses can be modeled in dynamical simulations via the last term in Eq.~\eqref{eq:eGPE}. For atomic gases, the dominant loss channel stems from three-body collisions, where two particles can form a dimer, while the third particle absorbs the excess energy and momentum. The corresponding cubic density dependence of the loss rate \cite{Moerdijk1996} is captured by the last term in Eq.~\eqref{eq:eGPE} with $\alpha=3$. The loss coefficient $L_3>0$ depends on the particle interactions \cite{Fedichev1996, Kohler2002, Ferrier-Barbut2013} and assumes typical values of $1.3\times10^{-41}\,\mathrm{m}^6\, \mathrm{s}^{-1}$ ~\cite{Schmitt2016} for ${}^{164}$Dy,
 $1.5\times10^{-40}\,\mathrm{m}^6\,\mathrm{s}^{-1}$ at $a_s=94a_0$~\cite{Boettcher2019} for ${}^{162}$Dy atoms. These numbers have significant influence on the achievable densities and observation times in experiments with atomic dipolar condensates.

 In the context of MW-shielding of molecules, the behavior of inelastic collisions differs significantly. As discussed in Sec.~\ref{sub:sec:three-body}, three-body loss primarily becomes significant in the regime where bound field-linked states are present. In the absence of field-linked bound states, inelastic two-body loss can increase due to stronger localization of the scattering wave function near the shielding barrier. This enhanced localization facilitates pre-dissociation into lower-lying dressed states or tunneling to short range. Such losses can be accounted for via the last term in Eq.~\eqref{eq:eGPE} with $\alpha=2$. In addition, one-body losses ($\alpha=1$) can arise from phase-noise-induced decoherence of the dressed states. Such losses stem from technical issues and can be suppressed in experiments, as we will discuss in more detail in Sec.~\ref{sub:sec:shieldingimplementation}.

\subsection{Mean-field stability and geometry}
\label{sub:sec:mf_stability}
For systems of dipolar atoms, stability on the mean-field level sets stringent requirements on confinement and interaction properties\cred{~\cite{Fischer2006}}, with the overall sum of all interactions in Eq.~(\ref{eq:eGPE}) always required to remain repulsive \cite{Lahaye2008,Koch2008}. Within the eGPE, the inclusion of quantum fluctuations stabilizes the condensate at a density that \cred{decreases with increasing interaction strength}. \cred{Interesting quantum phases typically emerge for $a_{\rm dd}>a_{\rm s}$. 
Realizing these phases requires sufficiently strong dipole-dipole interactions to maintain low densities and thereby suppress atomic three-body losses, which would otherwise lead to a dynamical collapse of the BEC, as observed in the first Cr BECs~\cite{Lahaye2008,Koch2008}. 
Such strong dipolar interactions have become possible with quantum gases of Er and Dy atoms.}

For dipolar molecules, the tunability of the contact interaction via field-linked resonances \cite{Lassabliere2018,Chen2023,Schmidt2022,Mukherjee2024} can therefore be a valuable asset to stabilize a quantum degenerate ensemble. Even far from resonance, $a_s$ can be tuned by a few hundred $a_0$ and sometimes features a zero crossing. With resonant static electric field shielding  \cite{Mukherjee2023b} the calculated contact interaction can be attractive in the entire regime of effective shielding. In such a case, strong confinement in a pancake-shaped geometry, with the electric dipoles polarized perpendicular to this pancake, might be required for dipolar repulsion to compensate for the strong contact mean-field attraction. 

\subsection{Droplets and supersolids on the mean-field level}
\label{sub:sec:DropSuso}

\cred{Supersolidity refers to a phase of matter that simultaneously exhibits superfluid transport and crystalline density order, implying the coexistence of off-diagonal long-range order and broken translational symmetry.} 

The study of structure formation, supersolidity and other quantum states in bosonic dipolar gases is historically motivated by the behavior of superfluid  $^4$He. Both helium and dipolar BECs exhibit a similar roton excitation spectrum, which introduces a natural length scale for structure formation and exotic superfluid properties~\cite{Landau41,Feynman57,Santos2003,Chomaz2018}. 

Under sufficiently high pressure, low-temperature helium forms a solid, in which a so-called supersolid phase has been speculated to exist \cite{Andreev1969,Chester1970,Leggett70,Gross58}. This supersolid phase is based on the appearance of defects in the from of vacancies or interstitials. If the energy gain from delocalization of defects is sufficient, they could exist in the ground state of an otherwise regular crystal and contribute to a finite superfluid flow \cite{Andreev1969,Chester1970,Leggett70}. Despite extensive efforts ~\cite{Kim2012,Balibar2010,Boninsegni2012}, the observation of defect-induced supersolidity has remained elusive to this day, and theoretical signatures could thus far only be detected in the ground state of bosons with soft-core repulsion~\cite{Cinti2014}.

In dipolar quantum gases, a different type of supersolid state can emerge in the form of a periodic density wave or a regular array of superfluid droplets, as originally described by \textcite{Gross58}. While weakly interacting, the formed superfluid array of quantum droplets -- each containing up to several hundred particles -- breaks continuous translational symmetry, exhibits phonon-like excitations, and thus shows the defining features of a supersolid phase. For ultracold atomic systems, the physics of such droplets and supersolids has been surveyed extensively in recent reviews~\cite{Lahaye2009,Boettcher2021,Chomaz2023,Mukherjee2023,Recati2023,Chomaz2025}. In the following, we highlight the key concepts that underpin droplet and supersolid formation in dipolar gases and discuss the potential of molecular BECs for further exploring their properties.  

In the absence of external confinement, the competition between the attractive part of the dipole-dipole interaction, short-range repulsion, and quantum fluctuations can give rise to the formation of stable self-bound solutions of the eGPE [see Eq.~\eqref{eq:eGPE}] \cite{Wachtler2016b,Chomaz2016,Saito2016,Baillie2016,Schmitt2016}. In the ground state of polarized dipoles, particles align along the polarization direction, minimizing the total energy and giving rise to an elongated droplet shape along the dipolar-axis. The application of some external harmonic confinement will exert an energy penalty that eventually leads to a splitting into multiple droplets beyond a critical trap frequency or critical particle number. If the droplets overlap sufficiently to maintain global phase coherence against quantum and thermal fluctuations, the system can support a finite superfluid flow, and thus forms a supersolid state. The phase transition to a density-modulated supersolid state has been studied under quasi two-dimensional \cite{Zhang2019,Ripley2023,Bland2022,Zhang2021,Hertkorn2021b} and one-dimensional \cite{Roccuzzo2019,Blakie2020b,SanchezBaena2024} confinement, using the eGPE of Eq.~\eqref{eq:eGPE}. Depending on the parameters, the transition can be of first or second order \cite{Zhang2019}. In the former case of a discontinuous first-order transition, the abrupt onset of strong density modulations often implies a direct transition from an unmodulated phase to an insulating droplet-array without global phase coherence between the droplets, with only a small parameter region in which global phase coherence can be maintained. The existence of a second-order transition line in quasi-1D \cite{Blakie2020b} and a critical point in quasi-2D \cite{Zhang2019} geometries, on the other hand, provides a substantial parameter regime in which supersolidity can occur and be maintained at finite temperature. Experiments have demonstrated such supersolid phases in finite ensembles of up to several $10^4$ magnetic atoms under quasi-1D \cite{Tanzi2019,Boettcher2019,Chomaz2019} and quasi-2D \cite{Norcia2021} confinement. These observations of dipolar supersolids in atomic condensates have led to broad investigations of their properties and excitations \cite{Santos2003,Chomaz2018,Guo2019,Hertkorn2019,Natale2019,Tanzi2019b,Blakie2020,Hertkorn2021,Hertkorn2021b,Hertkorn2021c,Schmidt2021,Tanzi2021,Poli2021,Poli2024,SanchezBaena2024,Sanchez-Baena2023,Bland2022,Norcia2022,Blakie2023,Buehler2023,Ilg2023,Sindik2024,Hertkorn2024,Biagioni2024,Blakie2025,yapa2025,Kirkby2025}.

\begin{figure}[t!]
    \centering
\includegraphics[width=\columnwidth]{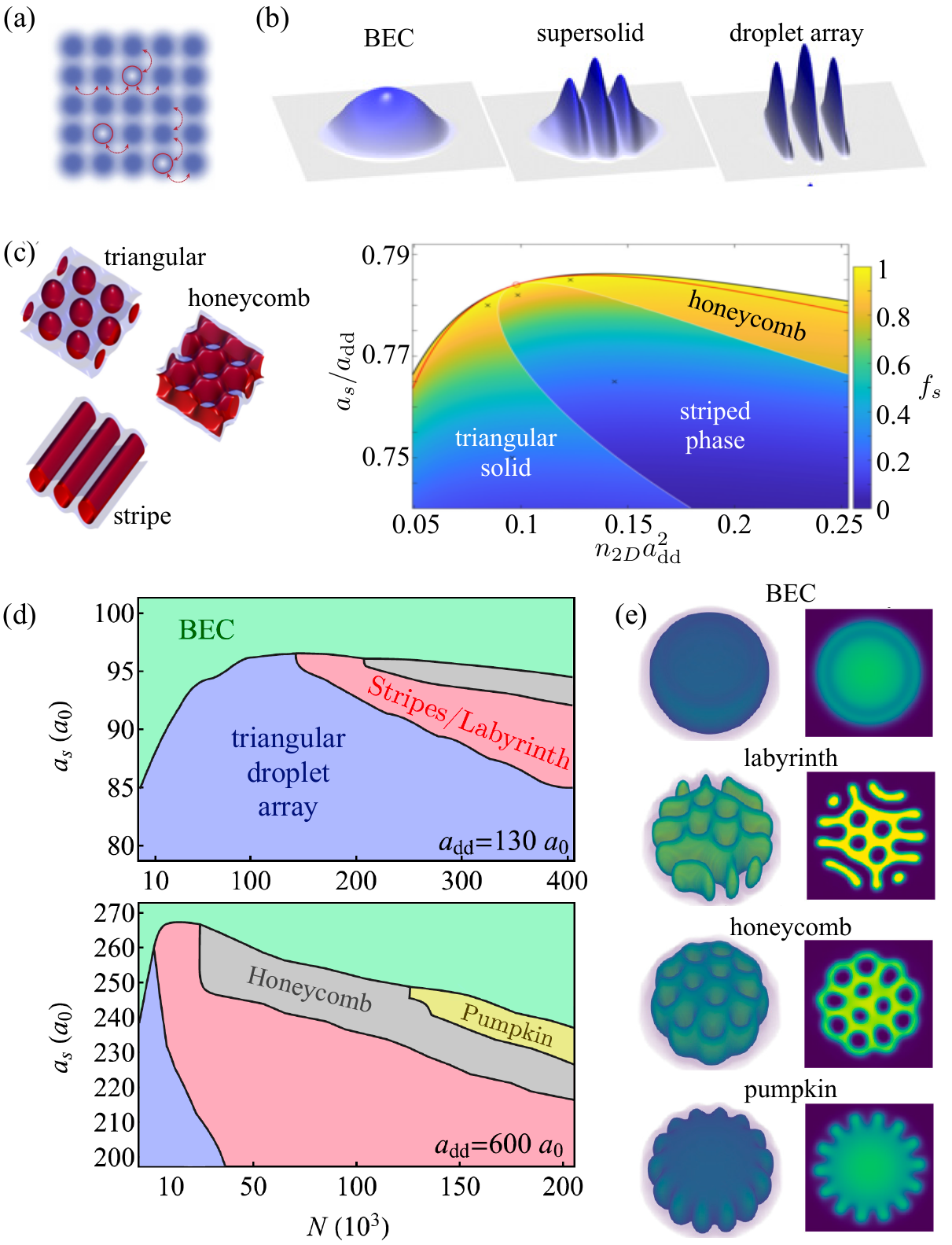}
    \caption{Supersolids and patterned states of dipolar quantum gases. (a) The concept of defect-induced supersolidity, as envisioned in helium, is based on the spontaneous formation of vacancies (red circles) which can generate a finite superfluid response of an otherwise perfect crystal. (b) In trapped dipolar gases dipolar interactions can induce the spontaneous formation of periodic density modulations that realize a supersolid phase, stabilized by quantum fluctuations. (c) In quasi-2D geometries such supersolid can emerge in the form of stripes, triangular lattices and honeycomb structures that can be explored by varying the different interaction strengths and the particle density. (d) In confined finite-size systems these underlying structures give rise to a rich phase diagram and (e) a broad spectrum of morphologies, that become accessible at lower particle numbers for the stronger interactions available with ultracold molecules. (a) Reproduced from \textcite{Balibar2010}. (b) Adapted from \textcite{Boettcher2021}. (c) Adapted from \textcite{Ripley2023}. (d) Adapted from \textcite{Schmidt2022}. (e) Adapted from \textcite{Hertkorn2021}.}
    \label{fig:dropletsandsupersolids}
\end{figure}

In two dimensions, the interplay between the dipolar mean-field and quantum fluctuations can give rise to more structures than simple triangular lattices, such as honeycomb structures \cite{Zhang2019}, stripe phases \cite{Hertkorn2021, Ripley2023}, and metastable ring-lattice phases \cite{Zhang2025c}, as illustrated in Figs.~\ref{fig:dropletsandsupersolids}(c) and (d). \cred{In finite confined systems, these thermodynamic phases—predicted in homogeneous systems and corresponding phase diagrams (e.g., \cite{Ripley2023})—are reflected as finite-size realizations in a rich spectrum of morphologies, ranging from droplet crystals, as well as stripe and labyrinth structures to pumpkin- or cogwheel-shaped states of finite-size condensates \cite{Hertkorn2021,Zhang2021}.} 

Probing these states in experiments with magnetic atoms requires particle numbers in excess of $10^5$ atoms for optimized trap geometries \cite{Hertkorn2021,Zhang2021} (see Fig.~\ref{fig:dropletsandsupersolids}). Such large atom numbers are challenging to achieve in magnetic BECs\cite{Norcia2021,Jin2023,Anich2024} and the associated densities are typically so large that three-body losses become significant during the structure-formation dynamics \cite{Zhang2021}.

This issue bears similarities to the mere stabilization of dipolar droplets, discussed above, which also requires strong dipoles that in atomic systems could only be achieved with Er and Dy quantum gases.   
\cred{In this context}, molecular gases with even stronger dipole-dipole interactions will therefore permit to probe the various structural transitions of dipolar supersolids at lower particle numbers and densities, ensuring sufficient stability against losses. This opportunity is illustrated in Figs.~\ref{fig:dropletsandsupersolids}(d) and (e), showing that a $4$-fold increase of $a_{\rm dd}$ decreases the particle number required to observe structural transitions by an order of magnitude. A more detailed discussion of the scaling behavior of trapped condensates with varying interaction strength can be found in \cite{Goral2000,Hertkorn2021,Hertkorn2021b}. In short, the eGPE in dimensionless units is parametrized by dimensionless contact and dipolar interaction strengths $\tilde{g}_\mathrm{s} \propto a_s N$, $\tilde{g}_\mathrm{dd} \propto a_\mathrm{dd} N$, elucidating that for larger $a_\mathrm{dd}$ a given $\tilde{g}_\mathrm{dd}$ is achieved for smaller $N$. Accordingly, the critical scattering length, at which a phase transition occurs ($\epsilon_\mathrm{dd} = \tilde{g}_\mathrm{dd}/\tilde{g}_\mathrm{s} \gtrsim 1$), shifts towards larger values. \cred{Additionally, the richer angular structure of the molecular interactions can lead to modifications in the droplet formation and shape~\cite{Baillie2025}.} Eventually towards stronger dipolar interactions, the underlying mean field theory in terms of the eGPE [see Eq.~\eqref{eq:eGPE}] approaches its limit of validity, and other approaches to treat strongly correlated system become necessary, which we will discuss in the following two sections.

\subsection{Limitations of mean-field theory}
\label{sub:sec:limitsofmeanfield}
A fundamental limitation of the mean field description based on the eGPE, given by Eq.~\eqref{eq:eGPE}, becomes clear upon inspecting the expression for the LHY energy correction 
\begin{align}\label{eq:LHY}
\gamma_{\rm qf}|\psi|^3=\sum_{\bf k}\tilde{V}({\bf k})\left(|v_{\bf k}|^2-u_{\bf k}v^*_{\bf k}\right)
\end{align}
that describes interactions between the condensate and excited Bogoliubov modes \cite{Schutzhold2006,Lima2011,Lima2012}. 
Here, $\tilde{V}({\bf k})$ is the Fourier transform of the molecular interaction potential. \cred{When} $\varepsilon_{\rm dd}>1$, the homogeneous BEC ($\psi={\rm const.}$) becomes unstable and gives way to supersolid patterns and quantum droplet solutions, as discussed in Sec.~\ref{sub:sec:DropSuso}, due to the softening of the Bogoliubov modes, $u_{\bf k}$ and $v_{\bf k}$. These soft modes generate complex contributions to the momentum sum in Eq.~\eqref{eq:LHY}, which become problematic in the LDA treatment of the LHY correction, used in the eGPE, Eq.~\eqref{eq:eGPE}. For the characteristic parameters of atomic BECs, the resulting imaginary part of $\gamma_\mathrm{qf}$ is typically small, and often neglected or eliminated via small-momentum cutoffs of the sum in Eq.~\eqref{eq:LHY} \cite{Wachtler2016,Bisset2016,Sanchez-Baena2023,He2024}, as mentioned above. Molecular condensates with stronger dipole-dipole interactions reach larger values of $\varepsilon_{\rm dd}$ for which the imaginary part of $\gamma_{\rm qf}$ becomes significant and even comparable to its real part \cite{Langen2025,Cardinale2025}. Resolving this inconsistency will require an improved theoretical framework for dilute but highly dipolar condensates, with a self-consistent inclusion of quantum fluctuations beyond leading-order LHY corrections and beyond the local-density approximation. 

The necessity to go beyond the common Hartree-Fock Bogoliubov framework \cite{Dalfovo1999} to account for quantum fluctuations  may also have implications for effects at finite temperature. Similar to quantum fluctuations, thermal fluctuations were found to be significant in atomic BECs, affecting the shape and phase boundary of quantum droplets \cite{Aybar2019} and even---counterintuitively---inducing the formation of supersolid crystals upon heating the system \cite{Sanchez-Baena2023,SanchezBaena2024,He2025}. Given the stronger interactions achievable in molecular BECs and hence larger fluctuations, understanding the effects of finite temperatures presents an important problem for ongoing and future work. Here, the softening of long-wavelength modes appears even more problematic than in the treatment of quantum fluctuations at zero temperature, because the thermal energy corrections comprise larger contributions from such excitation modes at longer wavelengths \cite{Sanchez-Baena2023,SanchezBaena2025}.  

Quantum and thermal fluctuations both lead to condensate depletion which becomes stronger for larger interactions. Recent calculations for MW-dressed molecular condensates predict a significant condensate depletion on the order of $20\%$ for $T=0$ and strong interactions \cite{Jin2025}, which is much higher than in typical atomic BECs~\cite{Lopez2017}. \cred{Microwave dressing comes with an effective hardcore repulsion and the angular anisotropy of the interaction potential can be modified compared to the standard dipole--dipole interaction~\cite{Baillie2025}. As a result, it remains an open question to what extent $\varepsilon_{\rm dd}$ alone provides a meaningful characterization of the interaction strength and the resulting condensate depletion in these systems.}

Perhaps even more strikingly, the electric DDI strength between molecules can reach values beyond the typical domain of dilute quantum gases, where mean-field theory becomes inapplicable. Strongly correlated phases, such as dipolar Wigner crystals \cite{Micheli2007}, may emerge at ultralow temperatures. Here, the vast tunability of molecular interactions, where the shielding barrier and potential minima can be controlled on the length scales of typical particle spacings, presents an exciting outlook for exploring exotic quantum phases in this strong-coupling regime.

\subsection{Beyond mean-field: Quantum Monte Carlo simulations}
\label{sub:sec:qmc}
Quantum Monte Carlo (QMC) methods \cite{Nightingale1998,Foulkes2001,Becca2017} offer a broad range of techniques for numerically exact studies of many-body problems. For the description of bosonic gases, path integral Monte Carlo (PIMC) sampling \cite{Ceperley1995, Boninsegni2006, Yan2017} provides an efficient and well developed approach to moderately-sized systems in thermal equilibrium. Continuous-space PIMC algorithms were developed to study strongly-correlated systems like superfluid or solid helium \cite{Ceperley1995}, and are, hence,  particularly well suited for simulations of molecular condensates. 

In the PIMC approach, the $N$-body density is approximated by its high-temperature limit, allowing one to evaluate the resulting high-dimensional path integral via Monte-Carlo sampling, using the so-called worm algorithms \cite{Boninsegni2006} for an efficient inclusion of the bosonic exchange symmetry. With advanced discretization schemes for 
such as the pair-product approximation \cite{Pollock1984, Cao1992, Yan2015} or fourth-order expansions \cite{Sakkos2009, Lindoy2018}  accurate descriptions of hundreds of strongly-interacting particles can be possible. 

\begin{figure}[tb!]
    \centering
\includegraphics[width=\columnwidth]{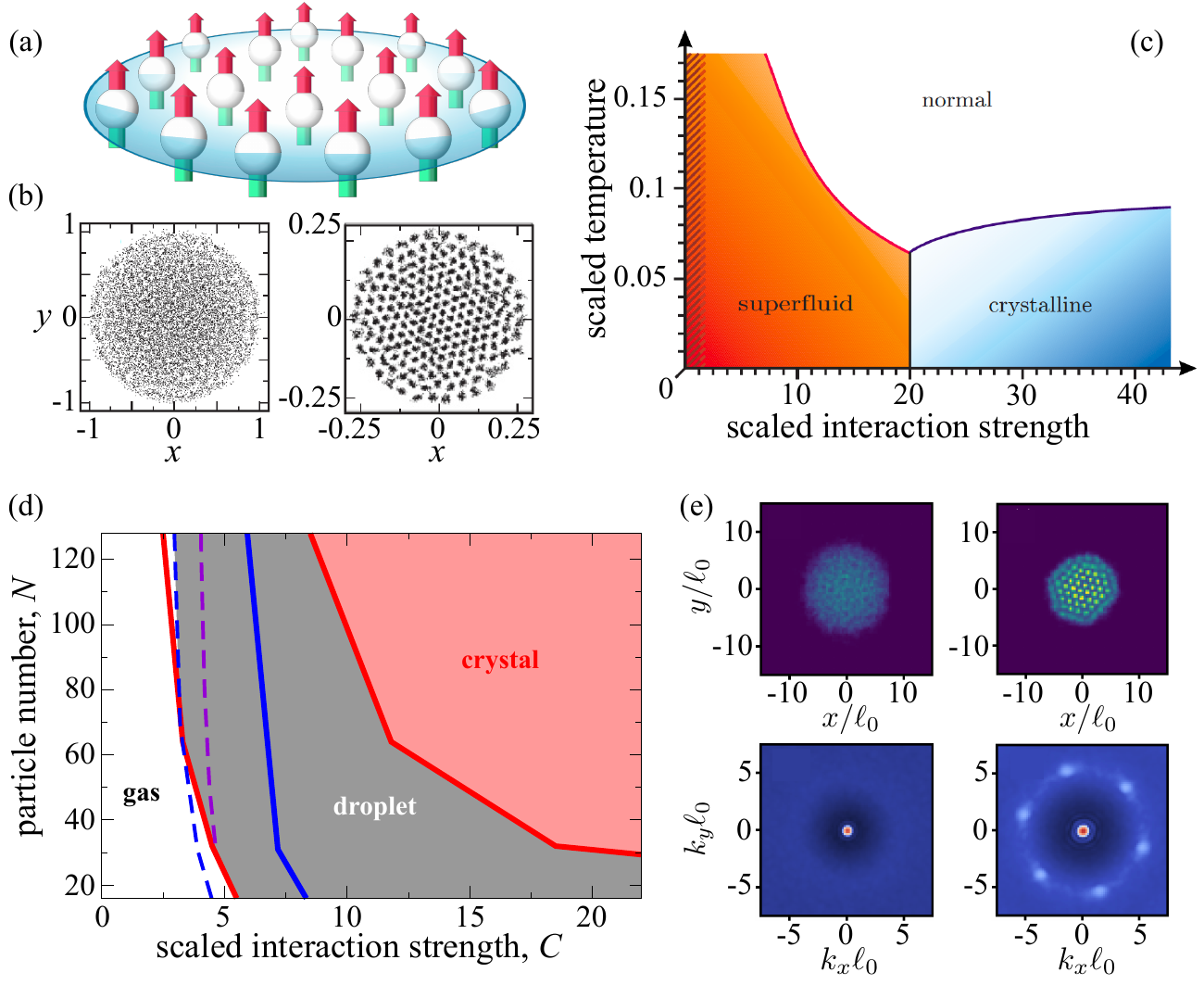}
    \caption{Strongly interacting dipolar gases.
    (a) A confined two-dimensional layer of repulsive dipoles can undergo Wigner crystallization, as illustrated in panel (b) for a harmonically confined finite system. Adapted from \textcite{Jain2011}. (c)~The corresponding thermodynamic phase diagram exhibits classical Wigner crystals as well as superfluid and normal fluid phases. Adapted from \textcite{Micheli2007}. (d) Microwave-dressed molecules, with interactions described by Eq.~\eqref{eq:Vpp1MW}, can form oblate self-confine quantum droplets, whose phase boundary has been determined with the eGPE (blue dashed) \cite{Langen2025}, variational calculations (purple dashed) \cite{Jin2025}, PIGS simulations (blue solid) \cite{Langen2025}, and PIMC simulations (red solid) \cite{Ciardi2025}. With increasing interactions, the droplet undergoes a phase transition to a two-dimensional superfluid membrane and eventually forms a two-dimensional self-confined molecular van der Waals crystal (solid red line), as illustrated in panel (e). Adapted from \textcite{Ciardi2025}.}
    \label{fig:beyondmeanfield}
\end{figure}

QMC simulations have been used to study strongly confined dipolar bosons in two spatial dimensions, where interactions are purely repulsive, as illustrated in Fig.~\ref{fig:beyondmeanfield}(a). Such simulations reveal Wigner crystallization at high densities \cite{Lozovik2004, Astrakharchik2007, Buechler2007, Mora2007, Jain2011, Macia2011} [see Fig.~\ref{fig:beyondmeanfield}(b,c)], which can feature stripe symmetry for tilted dipoles \cite{Macia2014, Bombin2017, Cinti2019, Bombin2019}. Ensembles of magnetic atoms have been simulated using an additional hard wall \cite{Nho2005, Saito2016} or a $V\propto r^{-6}$ or $V\propto r^{-12}$ repulsive potential to model atomic $s$-wave scattering. These simulations have confirmed the formation of quantum droplets \cite{Macia2016, Boettcher2019b, Boninsegni2021, Ghosh2024, Bombin2024a} and droplet arrays \cite{Cinti2017a, Cinti2017b, Kora2019}, similar to predictions by mean-field calculations, but existing also in other parameter regimes. Notably, in such highly anisotropic self-assembled configurations, special care must be taken to handle metastable states~\cite{Ciardi2025}. Also, in simulations of weakly interacting atomic condensates, QMC approaches often struggle to reach the required large particle numbers of several $10^3$ atoms for a direct and accurate description of typical experimental conditions, whereby the length scale disparity between the short-range atomic repulsion and large inter-particle spacings implies additional numerical challenges. Due to the stronger interactions and typically smaller particle numbers in molecular condensates, together with the accurate knowledge of the MW-controlled interaction potentials, PIMC approaches indeed offer a well suited and powerful approach to explore the equilibrium quantum phases of molecular BECs. 

Initial studies have explored the formation of self-bound droplets by MW-dressed condensates \cite{Langen2025,Zhang2025,Ciardi2025, Cardinale2025}. For a single circularly polarized microwave field, as discussed in Sec.~\ref{sub:sec:shielding}, and using the analytical expression for the resulting interaction potential, the phase boundary between the droplet state and the dilute unconfined gas phase solely depends on the scaled interaction strength, Eq.~(\ref{eq:C}), and the number of molecules. Fig.~\ref{fig:beyondmeanfield}(d) compares the calculated phase boundary obtained in these recent studies, obtained from mean-field theory \cite{Langen2025} (blue dashed line), a variational approach \cite{Jin2025} (purple dashed line), path-integral ground-state (PIGS) simulations \cite{Langen2025} (blue solid line) and PIMC (red solid line) simulations \cite{Ciardi2025}. While all current theoretical results show qualitatively similar behavior, the predicted phase boundaries differ quantitatively. The PIGS simulations \cite{Langen2025} yield the largest critical interaction strength $C$ for droplet formation, exceeding the results of PIMC simulations \cite{Ciardi2025} by about a factor of $2$. The Monte Carlo results of \cite{Ciardi2025} are, however, close to eGPE simulations reported in \cite{Langen2025}. Both predict somewhat lower critical interaction strengths than the variational result of \cite{Jin2025}. This is consistent with the expectation that variational theory provides an upper bound on the system's ground-state energy and should therefore tend to overestimate the critical interaction strength for droplet formation, in accordance with \textcite{Ciardi2025}. Clearly, further work and in particular accurate comparisons with experimental measurements will be necessary to better understand the detailed phenomenology of droplet formation by strongly interacting molecules. 

Considering the similarities between the physics of ultracold molecular quantum gases and low-temperature superfluid helium, the peculiar and widely tunable shape of molecular interactions under MW dressing suggest new opportunities to explore exotic many-body phases. Indeed, recent QMC simulations of molecules in anti-dipolar configurations \cite{Ciardi2025} predict a new transition from the droplet phase to a self-bound superfluid membrane. Upon further increase of the interaction strength, this self-confined two-dimensional superfluid, consisting of a single molecular layer, undergoes a crystallization transition to a self-confined molecular crystal, as illustrated in Fig.~\ref{fig:beyondmeanfield}.

These recent findings offer an encouraging outlook for future explorations of exotic phases of strongly interacting bosons, such as helium-like, defect-induced supersolid states that have been predicted more than 50 years ago~\cite{Andreev1969}, but have thus far remained elusive \cite{Balibar2010}. The combination of numerical simulations and experimental measurements of molecular BECs may also shed light on the physics of quantum fluctuations beyond LHY corrections and eGPE approaches and provide an understanding of patterned droplet phases under strong interactions, as recently observed in first experiments \cite{Zhang2025b}.

\section{Experimental path to BEC}
As outlined in the previous sections, molecular BECs offer a wide range of open questions for exploration. The first molecular BEC was realized with NaCs in 2023 \cite{Bigagli2024} followed by NaRb in 2025 \cite{Shi2025DAMOP} using similar techniques. In the following, we summarize the central advances that were necessary to achieve these results, addressing both the resolved and the present technical challenges.

\subsection{Implementation and quality of shielding}

\label{sub:sec:shieldingimplementation}
While we have discussed the physics of shielding techniques in Sec.~\ref{sec:few-body}, we now focus on their implementation. Such implementations are crucial for the efficient cooling of a molecular sample and need to be specifically chosen and adapted to the species and experimental situation at hand. 

For example, when molecules are confined in optical dipole traps that are either near resonant or have particularly high intensities, tensor light shifts can be present in the rotationally excited states. The polarization of the trapping light can then compete as a quantization axis with the magnetic and the shielding field \cite{Anderegg2021,Zhang2024}. To mitigate such effects, effective shielding requires sufficiently strong shielding fields to outweigh other couplings.

Generally, the techniques discussed in Sec.~\ref{sub:sec:shielding} for shielding and tuning interactions rely on relatively strong static or MW electric fields. Generating such fields with sufficient stability and tunability can be experimentally challenging---especially given the importance of preserving as much as possible the optical access available in the vacuum environments of ultracold-molecule experiments. Field inhomogeneities can degrade trap confinement \cite{Gempel2016,Covey2017,Lassabliere2018,Valtolina2020} or perturb molecular trajectories in time-of-flight measurements. In the following, we discuss in more detail the corresponding considerations and technical constraints relevant to both static and MW fields.

\paragraph{Static field shielding}
For both confinement dc shielding and resonant dc shielding, large static electric fields of many $\text{kV}/\text{cm}$  are required, which are challenging to realize. Static electric fields are typically applied with either electric field rods or with transparent glass plate electrodes coated with indium tin oxide (ITO) \cite{Gempel2016,Covey2017}. A thin ITO layer features high transparency in the visible and near-infrared regime and thereby provides good optical access. However, finite absorption limits the intensity of optical dipole traps that pass through ITO electrodes.

High-voltage electrodes in close proximity to insulator materials can cause charges emitted by an electrode to trigger secondary electron emission avalanches~ \cite{Ziemba2021} or radiation hazards~\cite{West2017}. Remaining charges on the insulator lead to unintended dc Stark shifts of the molecules, which impede the precise manipulation of the internal molecular states. Therefore, it is paramount to avoid field enhancement at the edges of electrodes and at surface imperfections \cite{Covey2017}. 

Stability of static electric fields on the parts-per-million level is needed for reliable internal state manipulation. This can be realized by state-of-the-art high-voltage servos \cite{Shaw2015,Covey2017} but comes at the cost of reducing the speed at which electric fields can be varied, which---in turn---limits the speed at which dipolar interactions can be tuned. 

\paragraph{MW shielding}

MW shielding circumvents a number of the technical challenges associated with dc shielding and implementation into existing experimental setups is relatively straightforward.

MW shielding itself still requires relatively strong microwave fields. For the MW-field generation various types of antennas can be used depending on the wavelength and accessibility of the experimental setup. So far, individual helical antennas \cite{Schindewolf2022}, helical antenna arrays \cite{Anderegg2021}, dual-feed waveguide antennas \cite{Lin2023,Chen2023}, and combinations of full-wave loop antennas \cite{Yuan2023,Bigagli2023,Shi2025DAMOP} have been used to generate circular polarized MWs, while individual loop antennas can add linear polarized MWs \cite{Bigagli2024,Shi2025DAMOP}.

However, waveguide and loop antennas have poor directionality. To realize high field strengths it is thus desirable to place the molecules in the near field of these antennas. Helical antennas that are operated in axial mode have, in contrast, enhanced directionality, which scales with the number of turns of their helix \cite{Kraus1949}. They are particularly well suited for \cred{short MW wavelengths corresponding to rotational transitions of molecules with large rotational constants} 
(e.g., CaF, AlF, YO). MW-field focusing through curved mirrors is another option if the wavelength is much shorter than the metallic openings and structures surrounding the molecules. \cred{At long wavelengths corresponding to molecules with smaller rotational constants} (e.g., RbCs, RbSr), the antenna designs discussed so far will likely become too large. In this case, high field strength might be achievable in the near field of electrodes that surround the molecules. In-vacuum MW cavities~\cite{Dunseith2015} could provide another way to maximize field strengths.

For MW shielding, the induced dipoles of the molecules are synced by the MW field. However, phase noise at detunings comparable to $\Omega_\text{eff}$ can couple the dressed states $\ket{+}$ and $\ket{-}$. This effectively leads to decoherence of the rotational superposition and results in unshielded inelastic collisions. While the inelastic loss mechanism is a two-body process, the time signature is dominated by one-body decoherence with a decay rate of $\Omega^2 S_\phi(\Omega_\text{eff})/2$, where $S_\phi(\Omega_\text{eff})$ is the single-sideband phase noise \cite{Anderegg2021,Chen2023PhD}. To reach one-body lifetimes of more than $10$\,s one typically requires $S_\phi(\Omega_\text{eff}) < -170\,\text{dBc}/\text{Hz}$. Given that state-of-the-art MW sources have typically phase-noise levels of $-160\,\text{dBc}/\text{Hz}$ at a detuning of $10$\,MHz and that Johnson--Nyquist noise is about $-174\,\text{dBm}/\text{Hz}$ at room temperature, it has become an established method to use MW bandpass filters in combination with low-gain high-power amplifiers as output stage \cite{Bigagli2024,Biswas2025,Lin2023}.

While less critical, amplitude noise can in principle also couple MW dressed states by modulating $\Omega_\text{eff}$ through $\Omega$. Finite detunings can suppress this coupling.

\textcite{Karman2025} studied the effect of typical imperfections of the MW polarization in the context of double MW shielding. The imperfections prevent full compensation of the dipolar interaction in double MW shielding. In the compensation regime, the effect of finite ellipticity dominates, while a typical finite tilt of the $\pi$ field with respect to the $\sigma$ field can be neglected.

\subsection{Making molecular BECs}

For a trapped gas in a 3D harmonic trap, Bose--Einstein condensation occurs when the phase-space density, $\rho = \lambda_{\rm dB}^3 n_0$, exceeds $\zeta(3) \approx 1.202$ \cite{Dalfovo1999, Ketterle1999}. Here, $\lambda_{\rm dB}$ is the thermal de Broglie wavelength in a gas and $n_0$ is the peak density. To meet this condition, temperatures need to reach the ultracold regime while maintaining sufficiently high particle densities. In general, molecules can be brought to ultracold temperatures ($\sim 1\,\mu$K) by association \cite{Ni2008,Danzl2008} or direct laser cooling \cite{Fitch2021,Langen2023}. BEC of molecules requires further cooling to temperatures in the 10-nK regime. Despite important progress in laser cooling, dipolar molecular BECs so far have been produced only by association. We briefly summarize the procedure in the following. 

To generate dense molecular gases in the $\mu$K regime, first clouds of two atomic species are laser cooled, sympathetically cooled, and then transferred into diatomic molecules by magneto-association \cite{Kohler2006,Chin2010}. In magneto-association, atom pairs are adiabatically transferred from the scattering continuum into a weakly bound molecular state via an  interspecies magnetic Feshbach resonance. The resulting molecules---so-called Feshbach molecules---typically have a binding energy of about 0.1--10\,MHz and a negligible permanent dipole moment. Feshbach molecules are then transferred to the ro-vibrational ground state through a coherent two-photon process known as stimulated Raman adiabatic passage (STIRAP) \cite{Stwalley2004,Vitanov2017}. The power of the association technique is that the resulting molecules inherit the cold temperature and high phase-space density from the atoms. The process results in ultracold gases of ground-state molecules with binding energies of 100s of THz and a significant permanent dipole moment, typically at temperatures of about 1\,$\mu$K.

Association from overlapping atomic BECs \cite{Warner2021, Lam2022, Wang2016} or from dual single-shell Mott insulators in an optical lattice \cite{Reichsollner2017} may appear natural to realize even colder molecular samples, but have so far been relatively inefficient. In practice, it has proven advantageous to associate molecules from near-degenerate thermal atomic mixtures, maximizing the number of molecules in the resulting samples, typically around $10^4$--$10^5$ molecules at a temperature of about 1\,$\mu$K, corresponding to a phase-space density of about $10^{-3}$ \cite{Schindewolf2022,Bigagli2023,Lin2023}. This approach has also been adopted by the NaCs and NaRb BEC experiments, relying on efficient evaporative cooling of the molecular gas, enabled by collisional shielding. In evaporative cooling, the hottest molecules in a thermal sample are selectively removed, and then, when the sample rethermalizes through collisions, the temperature will be reduced. 

The development of double MW shielding was key to realizing the first molecular BEC via evaporative cooling of NaCs molecules \cite{Bigagli2024,Karman2025,Deng2025}. This technique also enabled Bose--Einstein condensation of NaRb \cite{Shi2025DAMOP} and is expected to be adaptable to a broad range of molecules \cite{Karman2025}. Double MW shielding provides high suppression of inelastic losses, while maintaining sufficiently strong interactions for efficient thermalization.

\begin{figure}[tb]
    \centering    
    \includegraphics[width=0.485\textwidth]{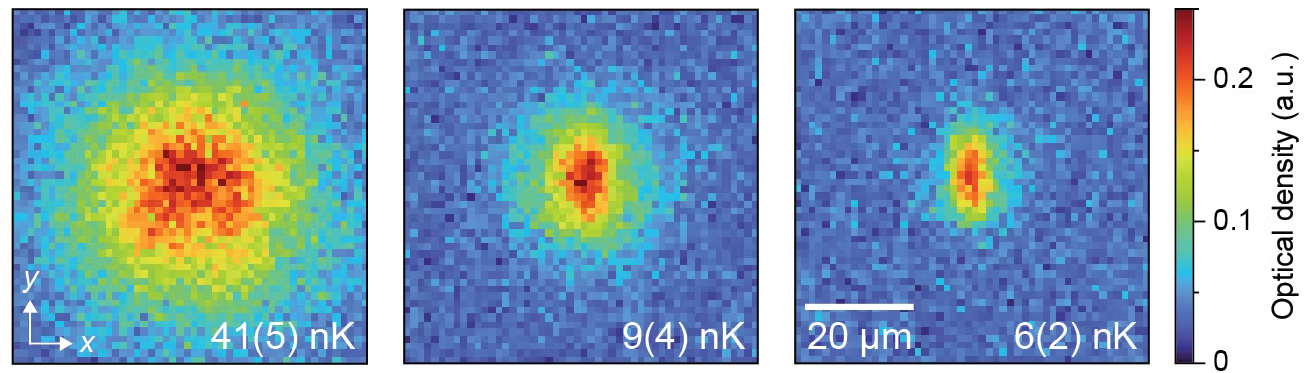}
    \caption{Creation of the first molecular BEC. Transition from a thermal gas to a BEC of NaCs molecules (left to right). Similar results have recently also been obtained for NaRb \cite{Shi2025DAMOP}. Adapted from~\textcite{Bigagli2024}.}
    \label{fig:BECresults}
\end{figure}

As quantum degeneracy is approached, the momentum distribution of an ultracold gas changes dramatically and develops a characteristic peak structure~\cite{Ketterle1999}. First, as the phase-space density approaches 1, the bosonic enhancement becomes significant and the profile changes from a Gaussian to a Bose-enhanced Gaussian. Once a BEC emerges, the momentum distribution becomes bimodal, with the molecules in the BEC occupying the interaction-broadened lowest momentum state in the trap, and the remaining thermal molecules follow a Bose-enhanced profile. As evaporation proceeds, the thermal fraction of molecules is reduced until a quasi-pure BEC remains~\cite{Bigagli2024,Zhang2025b}. Fig.~\ref{fig:BECresults} shows this evolution of the momentum distribution for the first molecular BEC. 

\subsection{Probing BECs}

The momentum distribution in these experiments is measured using absorption imaging after time-of-flight expansion \cite{Bigagli2024,Shi2025DAMOP}. Absorption imaging of molecules comes with additional challenges compared to atoms:

First, during expansion, the molecules must remain in the collisionally shielded state. The density in the BEC before expansion is so high that the entire BEC would be lost within about 1\,ms if unshielded. Only at the end of the expansion does the density become much lower, and this condition is relaxed. 

Second, for molecules like NaCs and NaRb, direct detection is not possible and the molecules must be dissociated into their atomic components for detection. It is important that dissociation happens rapidly at the end of time-of-flight expansion and without introducing additional energy to prevent a distortion of the measured momentum distribution, which would lead to inaccuracies in thermometry~\cite{Lam2022}. Reverse STIRAP from the molecular ground state directly into the continuum of unpaired atomic states fulfills these requirements. It allows dissociation within about $10$\,$\mu$s and does not add extra energy---in contrast to reverse STIRAP to a weakly bound molecular state, followed by magnetic dissociation. 

Lastly, the need for reverse STIRAP at the end of time-of-flight expansion requires the STIRAP lasers to hit the molecular cloud after a substantial amount of free fall (up to about $1\,\text{mm}$). This can be be achieved either by implementing vertical STIRAP beams \cite{Bigagli2023, Bigagli2024}, propagating in the direction of gravity, or by dynamically changing the pointing of horizontally propagating STIRAP beams during time of flight \cite{Shi2025DAMOP}. Also the shielding fields need to be very uniform.  Gradients could lead to a force on the cloud during time-of-flight expansion and push it out of the STIRAP beams; curvature could lead to a distortion of the momentum distribution. These experimental techniques have been shown to allow for detection efficiencies over $40 \%$ after 25\,ms of time-of-flight expansion \cite{Zhang2025b}.

The available detection techniques for a molecular BEC impact the range of physics that can be explored. Early experiments of atomic BECs relied almost exclusively on absorption imaging after time-of-flight expansion \cite{Ketterle1999}, naturally steering investigations toward phenomena accessible with this technique—such as collective excitations, superfluidity, interference, and thermodynamics. At this point, molecular BECs are similarly poised for studies of bulk properties, exploring dipolar quantum gases and quantum liquids with strong and highly tunable interactions. The recent observation of strongly interacting droplet phases in a BEC of NaCs molecules by \textcite{Zhang2025b} is a first example. We expect that detection methods can be further expanded to high-resolution \textit{in situ} imaging. The relevant techniques have been developed and are in wide use in the context of atomic quantum gases. They should be applicable with straightforward modifications to the detection of molecular quantum gases after reverse STIRAP, as discussed above~\cite{Rosenberg2022,Christakis2023,Mortlock2025}. They also appear to be even more readily implementable in laser-coolable molecules that can scatter many photons for efficient fluorescence detection~\cite{Anderegg2019}, although BECs of such species have not yet been realized. 

\subsection{Larger BECs}
As the creation of molecular BECs requires a complex multi-stage process---involving evaporative cooling of ultracold atomic gases, followed by molecular association, and evaporation of molecules---the sizes of molecular BECs are naturally smaller than typical atomic BECs. To date, NaCs BECs of up to 2000 molecules \cite{Bigagli2024,Zhang2025b} and NaRb BECs with about 600 molecules \cite{Shi2025DAMOP} have been reported, whereas atomic condensates routinely reach $10^5$ atoms. While already relatively small molecular BECs offer exciting scientific opportunities---both on the experimental and the theoretical side---an increase of BEC size in experiments will be highly desirable. 

The main avenues to reaching larger molecular condensates are (1) larger molecular samples at the beginning of evaporative cooling, and (2) molecular evaporation with higher efficiency.

Larger samples of ultracold molecules can be achieved by increasing the size of the initial atomic samples and a higher efficiency of magneto-association. However, it is not straightforward to reap the benefits: increasing the size of the initial atomic samples typically also raises the atomic density, which can reduce the pairing efficiency and even decrease the final number of molecules. This challenge can be mitigated by using atomic density profiles that are more uniform, for example, by reducing the geometric mean trap frequency in the harmonic trap, $\bar{\omega}$---as demonstrated for NaCs \cite{Lam2022}---or by employing box traps \cite{Navon2021}, as in the case of fermionic NaK \cite{Bause2021}. 

In order to further improve evaporative cooling of molecules, further improvement of the ratio of elastic (``good'') to inelastic (``bad'') collisions, $\gamma$, is highly desirable. For unshielded molecules, $\gamma \approx 1$, preventing efficient evaporative cooling. As discussed in Sec.~\ref{sec:few-body}, shielding both increases the elastic cross-section, $\sigma_{\rm el}$, and reduces the inelastic cross-section. For highly polar molecules such as NaCs and NaRb, shielded samples reside in the hydrodynamic regime, which occurs when a trapped gas undergoes many scattering events within one trap-oscillation period, i.e., when $\sigma_{\rm el}, n_0 > \bar{\omega}$ \cite{Ma2003,Schindewolf2022, Bigagli2023}. In this regime, the thermalization rate saturates at the geometric-mean trap frequency $\bar{\omega}$, and further increases in the elastic rate have little effect. This shifts the focus toward minimizing inelastic processes.

There are a number of inelastic channels that should be suppressed to achieve efficient evaporation and long-lived degenerate molecular gases:

First, off-resonant scattering of the trapping light \cite{Grimm2000}. Absorption of a single photon from the optical dipole trap can change the internal state of a molecule, resulting in loss. Even if the molecule remains trapped, it will typically be lost in a subsequent collision. This fundamental process generally limits the one-body lifetime of a collisionally shielded gas to approximately~$10\,$s.

Second, microwave phase noise, which can flip molecules into unshielded dressed states \cite{Anderegg2021,Schindewolf2022}. This process has been widely observed in recent shielding experiments \cite{Bigagli2024,Chen2023PhD,Biswas2025,Lin2023,Yuan2025,Shi2025DAMOP,Zhang2025b} and effectively mitigated via the use of narrow microwave filter cavities. Those cavities dramatically reduce phase noise, yielding lifetimes that exceed $50\,$s.

Third, two-body collisional loss. While shielding strongly suppresses two-body loss, some residual loss remains. It has been shown that the loss rate coefficients can be suppressed to $\beta_{\rm 2B} \approx 10^{-14}$ (10$^{-13}$)\,cm$^{3}$/s for NaCs (NaRb)~\cite{Yuan2025,Shi2025DAMOP}. At typical BEC densities of $10^{12}$\,cm$^{-3}$, this corresponds to lifetimes of about 100 (10)\,s. These numbers are already  impressive, corresponding to an improvement by at least three orders of magnitude compared to the unshielded case. Further improvements are predicted when using molecules with very large dipole moments, such as KAg or CsAg (see Tab.~\ref{tab:mwdressing}), where two-body loss can be suppressed by another one or two orders of magnitude~\cite{Karman2025}.

Fourth, three-body recombination~\cite{Stevenson2024}. The current understanding is that double MW shielding leads to three-body recombination rates that are consistent with zero~\cite{Yuan2025}, when used in the regime where field-linked bound states are absent.

So far, molecular evaporation efficiencies of $\alpha \approx 2.5$ have been reached \cite{Toprakci2025DAMOP, Zhang2025b}, which is close to evaporation efficiencies reached in typical atomic quantum gas experiments. Here, $\alpha = - \ln{ d \rho} / \ln{d N}$ and $\rho$ is the phase-space density. Thus, for every order of magnitude lost in number, the molecules gain around 2.5 orders of magnitude in phase-space density. In addition, laser-cooled molecules have now taken initial steps into the collisional regime, providing an additional platform for future exploration~\cite{Burau2024,Jorapur2024}. The very best atomic evaporation experiments achieve $\alpha = 4$ \cite{Olson2013}. If inelastic decay can be further reduced---which appears within experimental reach---and molecule experiments achieve $\alpha = 4$, this would increase the final particle numbers in BECs by an order of magnitude. Such improvements, together with advances in pairing efficiency and trap design, bring molecular BECs with more than $10^4$ molecules into realistic reach. 

At this point, the field of molecular BECs is moving from first proof-of-principle realizations to a regime in which strongly dipolar condensates can systematically address open questions on dipolar many-body physics and associated novel states of matter.

\section{Conclusion}

The realization of strongly dipolar molecular BECs marks a new era in the study of quantum matter with dipolar interactions. In these systems, the interplay between precisely engineered shielding schemes and tunable interactions enables the controlled exploration of dipolar interactions beyond the paradigms established with magnetic atoms. The successful creation of stable, shielded molecular condensates demonstrates that losses can be suppressed to the point where genuine many-body physics becomes accessible, opening the door to the further exploration of novel quantum phases such as dipolar supersolids and droplets.  Together with other recent developments such as the realization of quantum gas microscopy with molecules~\cite{Rosenberg2022,Christakis2023,Mortlock2025}, long-lived rotational coherence~\cite{Hepworth2025}, Floquet engineering of spin models~\cite{Miller2024}, quantum gates~\cite{Holland2023,Bao2023,Picard2025,Ruttley2025}, and schemes for the implementation of synthetic gauge fields~\cite{Xu2025b} and SU($N$) symmetry~\cite{Mukherjee2025b,Mukherjee2025c}, these advances establish molecular BECs as a uniquely versatile platform for probing fundamental questions in dipolar quantum matter, from few-body physics to the emergence of collective phenomena in systems dominated by anisotropic, long-range interactions.

\section*{Acknowledgments}
The authors thank Kasper Rønning Pedersen for valuable input on the pseudopotential interactions, and Xing-Yan Chen and Jing-Lun Li for helpful comments on the manuscript. We acknowledge discussions with Hans Peter B\"uchler, Tobias Ilg and Krzysztof Jachymski, as well as with the participants of the workshop ``New directions in cold and ultracold chemistry'' at the Lorentz Center Leiden. This work was performed in part at the workshop ``Cold and ultracold molecules for fundamental physics and many-body quantum science'' at Aspen Center for Physics, which is supported by National Science Foundation grant PHY-2210452. 
J.H.\ gratefully acknowledges support by the MIT Pappalardo Fellowships in Physics and financial support by the Vector Stiftung (project No.\ P2021-0114). 
T.K.\ acknowledges NWO VIDI (Grant ID 10.61686/AKJWK33335).
D.W.\ is supported by the Hong Kong RGC General Research Fund (Grants 14301620 and 14302722) and the Collaborative Research Fund (Grant No.\ C4050-23G).
S.W.\ acknowledges support from the Gordon and Betty Moore Foundation (Award No.\ GBMF12340). T.L.\ acknowledges funding from the European Research Council (ERC) under the European Union’s Horizon 2020 research and innovation program (Grant agreement No.\ 949431) and from Carl Zeiss Foundation. T.P. and T.L. acknowledge funding by the Austrian Science Fund (FWF) 10.55776/COE1 and the European Union (NextGen-
erationEU). 

\appendix

\section{Molecule parameters}
The internal dipole moment and the rotational constant in the lowest vibrational state of the electronic ground state are given in Table \ref{tab:const} for the various molecular species.

\newcolumntype{L}[1]{>{\raggedright\arraybackslash}p{#1}}
\begin{table*}[t]
\centering
\begin{ThreePartTable}
\caption{Parameters of relevant bosonic molecular species. The values are given for the electronic and vibrational ground state if not specified otherwise.}
\setlength{\tabcolsep}{5pt}
\begin{tabular}{l r@{}l r@{}l L{10.2cm}}
\hline
\hline
Molecule & \multicolumn{2}{c}{$d_0$ (D)} & \multicolumn{2}{c}{$B_\text{rot}/h$ (GHz)} & References\\
\hline
${}^7$Li${}^{23}$Na      &  0&.4513(4)   & 11&.220\,01(18) & \textcite{Graff1972,Fellows1991}\tnote{a}\\
${}^6$Li${}^{40}$K       &  3&.3992(15)  &  8&.742(3) & \textcite{Dagdigian1972,Bednarska1996,Yang2020}\tnote{a}\\
${}^7$Li${}^{85}$Rb      &  3&.9810(22)  &  6&.4695(9) & \textcite{Dagdigian1972,Dutta2011}\tnote{a}\\
${}^7$Li${}^{133}$Cs     &  5&.5(2)\tnote{b} &  5&.617\,43 & \textcite{Deiglmayr2010,Staanum2007}\\
${}^{23}$Na${}^{39}$K    &  2&.7358(14)  &  2&.848\,1659(5) & \textcite{Dagdigian1972,Yamada1992}\tnote{a}\\
${}^{23}$Na${}^{87}$Rb   &  3&.2(1)      &  2&.089\,6628(4) & \textcite{Guo2016,Guo2018}\\
${}^{23}$Na${}^{133}$Cs  &  4&.55(3)     &  1&.735\,616 & \textcite{Dagdigian1972,Docenko2004,Stevenson2023}\tnote{a}\\
${}^{41}$K${}^{87}$Rb    &  0&.574(17)   &  1&.095\,362(5)\tnote{c} & \textcite{Ospelkaus2010b,Ni2009}\\
${}^{39}$K${}^{133}$Cs   &  1&.86        &  0&.912\,68(18) & \textcite{Ladjimi2024,Zamarski2025}\\
${}^{87}$Rb${}^{133}$Cs  &  1&.225(3)(8) &  0&.490\,173\,994(45) & \textcite{Molony2014,Gregory2016}\\
${}^{39}$K${}^{107}$Ag   &  8&.50        &  2&.000\tnote{d} & \textcite{Smialkowski2021}\\
${}^{87}$Rb${}^{107}$Ag  &  9&.03        &  1&.092\tnote{cd} & \textcite{Smialkowski2021}\\
${}^{133}$Cs${}^{107}$Ag &  9&.75        &  0&.807\tnote{d} & \textcite{Smialkowski2021}\\
${}^{87}$Rb${}^{88}$Sr   &  1&.50        &  0&.548\tnote{cd} & \textcite{Ladjimi2024}\\
${}^{6}$Li${}^{53}$Cr    &  3&.3(2)      & 14&.02(4) & \textcite{Finelli2024}\\
${}^{40}$Ca${}^{19}$F    &  3&.07(7)     & 10&.267\,5373(3) & \textcite{Childs1984,Kaledin1999}\\
${}^{88}$Sr${}^{19}$F    &  3&.4963(6)(35) & 7&.487\,5989(8) & \textcite{Ernst1985,Colarusso1996}\\
${}^{138}$Ba${}^{19}$F   &  3&.1702(15)(32) & 6&.473\,958\,42(24) & \textcite{Ernst1986,Preston2025}\\
${}^{174}$Yb${}^{19}$F   &  3&.91(4)     &  7&.233\,8007(10) & \textcite{Sauer1996}\\
${}^{27}$Al${}^{19}$F    &  1&.515(4)    & 16&.488\,3548(3) & \textcite{Truppe2019}\\
${}^{89}$Y${}^{16}$O     &  4&.524(7)    & 11&.631\,900(6) & \textcite{Hoeft1993,Suenram1990}\\
\hline
\hline
\end{tabular}
\begin{tablenotes}
  \item[a] $d_0$ is derived using $B_0$ from one of the more recent publications.
  \item[b] in the second excited vibrational state.
  \item[c] derived from another isotopologue through the ratio of reduced masses.
  \item[d] at the equilibrium bond length.
\end{tablenotes}
\label{tab:const}
\end{ThreePartTable}
\end{table*}

\bibliography{biblio}

\end{document}